\documentclass[11pt,a4paper,oneside,notitlepage,onecolumn]{article}
\usepackage{amssymb}

\input{tcilatex}

\begin{document}

\title{Supernova Neutrino Spectrum with Matter and Spin Flavor Precession
Effects}
\author{\textbf{A. Ahriche\ \&}\ \textbf{J. Mimouni} \\
%EndAName
\textit{Physics department , Mentouri University, Route de Ain El-Bey,}\\
\textit{Constantine 25000, Algeria.}\\
$E-mail:${\small \ }$amin\_ahriche@hotmail.com,${\small \ }$jamalm@wissal.dz$%
}
\maketitle

\begin{abstract}
We consider Majorana neutrino conversions inside supernovae by taking into
account both flavor mixing and the neutrino magnetic moment. We study the
adiabaticity of various possible transitions between the neutrino states for
both normal and inverted hierarchy within the various solar neutrino problem
solutions. From the final mass spectrum within diffrent scenarios, we infer
the consequences of the various conversion effects on the neutronization
peak, the nature of final spectra, and the possible Earth matter effect on
the final fluxes. This enable us to check the sensibility of the \textit{SN}
neutrino flux on magnetic moment interaction, and narrow down possible
scenarios which depend on: the mass spectrum normal or inverted, the
solution of the solar neutrino problem; and the value of $\mu B$.
\end{abstract}

\section{Introduction}

The neutrino signal detected on the Earth from the \textit{SN1987A}
explosion \cite{SN1987} has opened up new ways to probe the neutrino
properties. Initially only constraints on the static properties of neutrino
during its interstellar journey (namely from vacuum oscillations) could be
acquired. Now with the entry into service of the large neutrino underground
detectors like Superkamiokande (SK) \cite{SK} and the Sudbury Neutrino
Observatory (SNO) \cite{SNO}, modification of the flavor content of the
neutrinos emerging from the \textit{SN} itself could be contemplated. This
is helped greatly by the new quantitative progress achieved in the past few
years in our knowledge of the mass squared differences and mixing parameters
from the atmospheric neutrino oscillations (SK and MACRO \cite{Macro}\ in
particular), and the solar neutrino oscillations \cite{Solarnu}.

Various works have appeared studying the consequences of both the matter
resonance effect (the so called \textit{MSW}\ effect) \cite{MSW}, and the
spin flavor precession (the \textit{RSFP} effect) \cite{RSFP}, especially
with the present data on the neutrino magnetic moment \cite{magnetic}, due
to the very large magnetic field in the pre-supernova interior \cite{RSFP+SN}%
. These studies have been mostly confined to the case of two active
neutrinos. The case of sterile neutrinos is no more rigorously pursued in
the light of the strong constraints on their existence from the atmospheric
neutrinos experiments. They used to be invoked in neutrino conversion, in
particular for their potential role in enabling r-process nucleosynthesis 
\cite{Nucl}.

We wish in this paper to consider the general case of three active neutrinos
on the neutrino \textit{SN} spectrum, namely the ones associated with the
three known leptonic flavors, in the context of both the \textit{MSW} effect
and the \textit{RSFP} effect. We have followed the seminal work of Dighe and
Smirnov \cite{DigheSmirnov} as far as their general line of attack on the
question, but generalizing it to take into account both of the above
mentioned effects at the same time. Considering the various schemes of
neutrino mass and mixing allowed by the data, we systematize the discussion
of the various spectrum distortion effects, and thus check the
discriminating power of such studies and how they could help resolve the
ambiguities associated with it, notably:

$\bullet $ the solar neutrino solution (which one to choose?)\footnote{%
The recent results from the KamLAND experiment \cite{Kamland}\ indicate that
the \textit{LOW} scenario is most probably ruled out, which leaves the 
\textit{LMA} as the most favored one.}.

$\bullet $ the type of hierarchy for the neutrino masses (\textit{normal} or 
\textit{inverted}).

$\bullet $ the value of $\mu B_{\perp }$ where $\mu $ is the\ neutrino
magnetic moment, and $B_{\perp }$ the \textit{SN} transverse magnetic field.

It follows from the existing solar and atmospheric neutrino data that
neutrino mass-squared differences satisfy the hierarchy $\Delta
m_{21}^{2}\ll \Delta m_{32}^{2}$, which permits us to order the three
mass-squared according to the two following cases: (\textit{i}) normal (or
direct) hierarchy, where $m_{1}^{2}\leq m_{2}^{2}\ll m_{3}^{2}$, and thus $%
\Delta m_{32}^{2}$\ will be positive (See Fig.1-a). (\textit{ii}) the
inverted mass hierarchy, where $m_{3}^{2}\ll m_{1}^{2}\leq m_{2}^{2}$ and
thus $\Delta m_{32}^{2}$\ will be negative (See Fig.1-b). We will assume
that $\Delta m_{21}^{2}$ is relevant for the solar neutrinos oscillations,
while $\Delta m_{32}^{2}$ is relevant for the oscillations of the
atmospheric neutrinos.

The effects of the neutrino conversions can be observed through, (\textit{i}%
) the disappearance (partial or complete) of the neutronization peak; (%
\textit{ii}) the interchange of original spectra and the appearance of a
hard $\nu _{e}$ spectrum; (\textit{iii}) the modification of the $\bar{\nu}%
_{e}$ spectrum; (\textit{iv}) the Earth matter effect, which is studied
taking into account neutrino mass and mixing in \cite{DigheSmirnov}. Note
that for significantly cosmological mass-squared differences ($\Delta
m^{2}=1\sim 100$ $\unit{eV}^{2}$), the spin-flavor precession and resonant
spin-flavor conversions may affect the supernova shock reheating and
r-process nucleosynthesis \cite{RSFP+SN}.

Although the ambiguities on the neutrino spectrum could not be solved, a
systematic study of their effects in the general three-generation case pave
the way for further constraining the various scenarios as new limits are
obtained from the detectors on the Earth.

\begin{center}
\bigskip \FRAME{ftbpFU}{14.1243cm}{4.9138cm}{0pt}{\Qcb{Neutrino mass
pattern: a) represents the case within normal mass hierarchy; and b)
represents the case within inverted mass hierarchy.}}{}{fig1ab.jpg}{\special%
{language "Scientific Word";type "GRAPHIC";maintain-aspect-ratio
TRUE;display "USEDEF";valid_file "F";width 14.1243cm;height 4.9138cm;depth
0pt;original-width 19.5837in;original-height 6.7499in;cropleft "0";croptop
"1";cropright "1";cropbottom "0";filename
'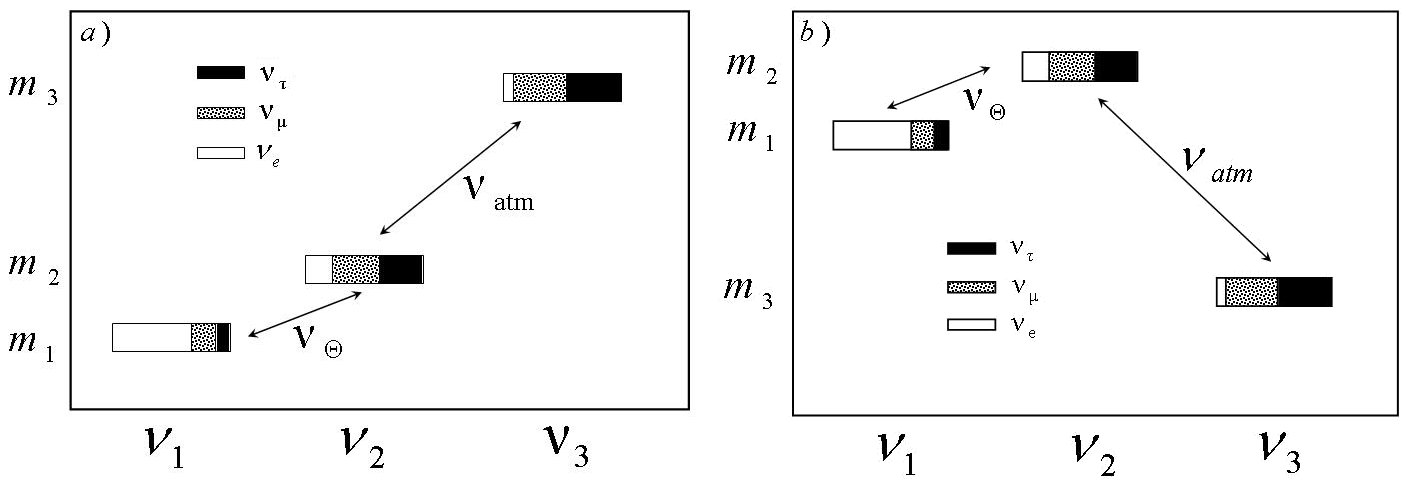';file-properties "XNPEU";}}
\end{center}

This paper is organized as follows: we first obtain the original neutrino
fluxes emerging from the \textit{SN} core; we then study neutrino conversion
outside the core for both normal and inverted mass hierarchy, and finally
determinate the final neutrino fluxes which reach the Earth detectors. The
various neutrino conversion effects are then classified according to the
different parameters relevant for the Solar neutrino solutions and the
magnitude of the magnetic interaction term $\mu B_{\perp }$.

\section{The Original Spectrum}

When the inner iron core of a massive star going through a type \textit{II}
supernova explosion becomes unable to support the electron degeneracy
pressure, most of the gravitational binding energy is released in the form
of a violent burst of neutrinos of all species \cite{Bethe,Suzuki}.

These neutrinos produced both during the neutronization burst and the
subsequent thermal cooling could undergo transformation of kind both inside
the \textit{SN}, and outside on their way to the Earth. We are however
interested only in their transformation within the \textit{SN}. In some
scenarios these transformations could boost the energy deposition at the
stalled explosion front as well as play a key role in the explosive
nucleosynthesis \cite{BetheWilson}. Yet, since in our study we are using
mass differences relevant to the Solar Neutrino Problem (\textit{SNP}) and
the Atmospheric Neutrino Problem only, these neutrinos transformations are
not expected to change the dynamics of the explosion as the corresponding
resonances take place outside the core.

Two effects have been widely considered: the first one is the matter effect, 
\textit{i.e.} the interaction of $\nu $'s with different matter constituents
which are mainly electrons, protons and neutrons. The second one consists of
the interaction of the neutrino's magnetic moment with the transverse
magnetic field of the \textit{SN}. For Dirac neutrinos, they flip into a
right handed $\nu _{R}$ kind which is known to be sterile since undetected
till now. Comparing the inferred energy output of \textit{SN1987A} to the
theoretical expectations, one can put strong constraints on these Dirac
neutrino conversions. On the other hand, when including the most general
mass terms in the simplest extension of the Weinberg-Salam model, the
neutrino fields, once diagonalized, turn out to be of Majorana kind. All
this concurs to make the Majorana neutrinos more palatable than the Dirac
kind in the context of \textit{SN} studies. We will assume in what follows
that the neutrinos are Majorana particles. Now since they are their own
antiparticles, only flavor conversion will be allowed.

In order to study these effects together, we should identify the profile of
both the density and the \textit{SN} magnetic field. We take the standard
parametrization: 
\begin{equation}
B(r)=B_{\perp o}\left( \tfrac{r}{r_{o}}\right) ^{-k};\text{ }r\geq r_{o}
\end{equation}
where $B_{\perp o}$ is the magnetic field strength at the distance \textit{r}%
$_{o}$\textit{=10}$\unit{km}$; and \textit{k=2} or \textit{3}; in our work,
we will consider $k=3$. The value of $B_{\perp o}$ lies between\footnote{%
This is taken to be the transversal value of the magnetic field strength at
the surface of the neutron star.} $\left( 10^{12}\sim 10^{15}\right) $%
\textit{G }, where \textit{G} is the strength magnetic field unit. For this
density profile, the important effects occur between\footnote{%
The resonance densities are approximately given by $\sim \frac{m_{N}\Delta
m^{2}\cos 2\theta }{2\sqrt{2}G_{F}E}\frac{1}{q}$, where $q$ is \textit{Y}$_{%
\text{\textit{e}}}$ for the \textit{MSW} transitions,\textit{\ }and \textit{%
1-2Y}$_{\text{\textit{e}}}$ for the \textit{RSFP} ones. The values of $%
\Delta m^{2}$ and $\theta $ are taken from the solar and atmospheric
neutrino data. \textit{Y}$_{\text{\textit{e}}}$ is generally set to be half,
and therefore \textit{1-2Y}$_{\text{\textit{e}}}$ lies between \textit{10}$^{%
\text{\textit{-4}}}$ to \textit{10}$^{\text{\textit{-3}}}$ \cite{WoosWea},
then all transitions occur in a range extending from a few $\unit{g}.\unit{cm%
}^{-3}$\ to \textit{10}$^{\text{\textit{7}}}\unit{g}.\unit{cm}^{-3}$.} $O$(1$%
\unit{g}\unit{cm}^{-3}$) and $O$(10$^{7}\unit{g}\unit{cm}^{-3}$), \textit{%
i.e.} this region may not be affected by the \textit{SN} shock wave. We will
thus take the progenitor profile for $\rho \gtrsim 1\unit{g}\unit{cm}^{-3}$
as approximately given by \cite{density} 
\begin{equation}
\rho Y_{e}\approx \rho _{o}\left( \tfrac{r}{r_{o}}\right) ^{-3}
\end{equation}
where $\rho _{o}=$\textit{2}$\times $\textit{10}$^{13}\unit{g}\unit{cm}^{-3}$%
. Of course, the exact shape of the density profile will depend on the
precise composition of the star.

The first question to answer is the flavor content of this burst when
reaching the conversion regions. The key point here is that they are
produced in a high density medium so that their mass eigenstates can readily
be inferred.

For this matter, let us rewrite the Hamiltonian in the flavored basis $\nu
^{f}$=($\nu _{e}$,$\nu _{\mu }$,$\nu _{\tau }$,$\bar{\nu}_{e}$,$\bar{\nu}%
_{\mu }$,$\bar{\nu}_{\tau }$)\ which is related to the mass eigenstates
basis in vacuum $\nu ^{v}$=($\nu _{1}$,$\nu _{2}$,$\nu _{3}$,$\bar{\nu}_{1}$,%
$\bar{\nu}_{2}$,$\bar{\nu}_{3}$) by the unitary transformation $\nu
^{f}=U\nu ^{v}$. We can then deduce the initial mass spectrum from the
evolution of neutrinos at very high density ($\rho \gtrsim 10^{8\text{ }}%
\unit{g}\unit{cm}^{-3}$).

Taking into account the neutrino effective potential coming from both the 
\textit{MSW} effect and the \textit{RSFP} effect, we can cast the
Hamiltonian into the following form: 
\begin{equation}
H=\text{Vacuum term}+\text{Matter term}+\text{Spin Precession term}
\end{equation}%
where each term, after the substraction of terms proportional to the unity,
is given by: 
\begin{equation}
\begin{array}{c}
\text{Vacuum term}=\frac{1}{2E}\left( 
\begin{array}{cc}
U & 0 \\ 
0 & U%
\end{array}%
\right) \left( 
\begin{array}{cc}
M^{2} & 0 \\ 
0 & M^{2}%
\end{array}%
\right) \left( 
\begin{array}{cc}
U^{\dagger } & 0 \\ 
0 & U^{\dagger }%
\end{array}%
\right) \\ 
\text{Matter term}=\alpha \rho \times \text{diag}\left\{ Y_{e}\text{,0,0,1-2}%
Y_{e}\text{,1-}Y_{e}\text{,1-}Y_{e}\right\} \\ 
\text{Spin Precession term}=\left( 
\begin{array}{cc}
0 & C. \\ 
-C & 0%
\end{array}%
\right) \text{, where }C=\left( 
\begin{array}{ccc}
0 & \mu _{e\mu } & \mu _{e\tau } \\ 
-\mu _{e\mu } & 0 & \mu _{\mu \tau } \\ 
-\mu _{e\tau } & -\mu _{\mu \tau } & 0%
\end{array}%
\right) \times B_{\bot }%
\end{array}
\label{H=}
\end{equation}%
with $M^{2}=$\textit{diag}$\left\{ 0\text{, }\Delta m_{21}^{2}\text{, }%
\Delta m_{31}^{2}\right\} $, $\alpha =\sqrt{2}G_{F}m_{N}$, $Y_{e}=\frac{n_{e}%
}{n_{e}+n_{n}}$ is the electronic fraction, the $\Delta m^{2}$'s\ are the
difference between squared masses; $\mu _{e\mu }$ is the transition magnetic
moment between the two flavored states $\nu _{e}\ $and $\nu _{\mu }$; and $U$
is the mixing matrix given by \cite{Maki} 
\begin{equation}
\left( 
\begin{array}{ccc}
c_{12}c_{13} & s_{12}c_{13} & s_{13} \\ 
-s_{12}c_{23}-c_{12}s_{23}s_{13} & c_{12}c_{23}-s_{12}s_{23}s_{13} & 
s_{23}c_{13} \\ 
s_{12}s_{23}-c_{12}c_{23}s_{13} & -c_{12}s_{23}-s_{12}c_{23}s_{13} & 
c_{23}c_{13}%
\end{array}%
\right)
\end{equation}%
where $s_{ij}=\sin \theta _{ij}$,$c_{ij}=\cos \theta _{ij}$ for $i$,$j=1$,$2$%
,$3$ ($i<j$); and $\theta _{ij}$\ is the vacuum mixing angle between $\nu
_{i}$\ and $\nu _{j}$. The mixing matrix terms satisfy the unitarity
relations: $\tsum\nolimits_{i}\left\vert U_{il}\right\vert
^{2}=\tsum\nolimits_{l}\left\vert U_{il}\right\vert ^{2}=1$.\ The equality
in the second term in Eq. (\ref{H=}), however, is only true to leading order
since radiative corrections induce tiny differences between the neutral
current potentials of $\nu _{e}$,$\ \nu _{\mu }\ $and $\nu _{\tau }$, and in
particular, results in a very small $\nu _{\mu }$-$\nu _{\tau }$\ potential
difference $\left\vert V_{\mu }-V_{\tau }\right\vert \sim $($10^{-5}\sim
10^{-4}$)$\times \alpha \rho $ \cite{Botella}.

Performing a rotation in the non-electronic subspace as ($\nu _{\mu }$,$\nu
_{\tau })\longrightarrow (\nu _{\mu ^{\prime }}$,$\nu _{\tau ^{\prime }})$
which diagonalizes the vacuum term and leaves the matter term invariant,
while for the magnetic term, the two values $\mu _{e\mu }$\ and $\mu _{e\tau
}$\ become nearly maximally mixed, and so whatever their values, the rotated
ones will be of the same magnitude, and therefore we have $\mu _{e\mu
^{\prime }}\sim \mu _{e\tau ^{\prime }}\sim \mu $. As for the third one $\mu
_{\mu \tau }$\ it remains invariant. Since $\nu _{\mu }$ and $\nu _{\tau }$\
are both produced via neutral currents only and are indistinguishable as far
as their detection, we take the non-electronic original fluxes to be
equivalent, so we can then write $F_{\mu }^{o}=F_{\tau }^{o}=F_{\mu ^{\prime
}}^{o}=F_{\tau ^{\prime }}^{o}$.

Comparing at very high density, the values of the various Hamiltonian terms
; we notice that for energies in the range of $\unit{MeV}$, $\alpha \rho \gg 
\frac{\Delta m_{21}^{2}}{2E}$,$\frac{\Delta m_{31}^{2}}{2E}$. Since we took
the term $\mu $\ to be between\cite{MagMoment} $10^{-12}\mu _{B}$ and $%
10^{-10}\mu _{B}$, the same thing applies when we compare the matter term
with the magnetic one with respect to the chosen range of $\mu B_{\perp o}$ 
\textit{i.e.} $\alpha \rho \gg \mu B_{\perp o}$. Note that the chosen values
of both $\mu $ and $B_{\bot o}$ don't affect the \textit{SNP} solution. We
then deduce that at the \textit{SN} core, the matter term is the dominant
one, and therefore the Hamiltonian is approximately diagonal 
\begin{equation}
H\simeq \alpha \rho \times \text{\textit{diag}}\left\{ \text{\textit{Y}}_{%
\text{\textit{e}}}\text{\textit{,0,0,1-2Y}}_{\text{\textit{e}}}\text{\textit{%
,1-Y}}_{\text{\textit{e}}}\text{\textit{,1-Y}}_{\text{\textit{e}}}\right\}
\end{equation}
This means that the matter eigenstates coincide with the flavored states,
and so, the neutrinos emerging from the \textit{SN} core are as follow:

\begin{tabular}{cccccc}
\textit{1- Normal Mass Hierarchy} &  & $\widetilde{\nu}_{1}^{o}\sim\nu
_{\mu^{\prime}}$ &  &  & $\widetilde{\bar{\nu}}_{1}^{o}\sim\bar{\nu}_{e}$ \\ 
&  & $\widetilde{\nu}_{2}^{o}\sim\nu_{\tau^{\prime}}$ &  &  & $\widetilde {%
\bar{\nu}}_{2}^{o}\sim\bar{\nu}_{\mu^{\prime}}$ \\ 
&  & $\widetilde{\nu}_{3}^{o}\sim\nu_{e}$ &  &  & $\widetilde{\bar{\nu}}%
_{3}^{o}\sim\bar{\nu}_{\tau^{\prime}}$%
\end{tabular}

\begin{tabular}{cccccc}
\textit{2- Inverted Mass Hierarchy} &  & $\widetilde{\nu }_{1}^{o}\sim \nu
_{\mu ^{\prime }}$ &  &  & $\widetilde{\bar{\nu}}_{1}^{o}\sim \bar{\nu}%
_{\tau ^{\prime }}$ \\ 
&  & $\widetilde{\nu }_{2}^{o}\sim \nu _{e}$ &  &  & $\widetilde{\bar{\nu}}%
_{2}^{o}\sim \bar{\nu}_{\mu ^{\prime }}$ \\ 
&  & $\widetilde{\nu }_{3}^{o}\sim \nu _{\tau ^{\prime }}$ &  &  & $%
\widetilde{\bar{\nu}}_{3}^{o}\sim \bar{\nu}_{e}$%
\end{tabular}

\section{The Neutrino Flavor Dynamics}

In order to find the final spectrum, we should study all possible
transitions between the neutrino species. We note that these transitions
occur in the \textit{isotopically neutral region}\footnote{%
The region characterized by \textit{1-2Ye=10}$^{\mathit{-4}}\mathit{\sim }$%
\textit{10}$^{\mathit{-3}}$, is called \textit{the isotopically neutral
region}.}, which consists mainly of layers of $^{4}He$, $^{12}C$, $^{16}O$, $%
^{28}Si$ and $^{32}S$ whose nucleus have $N\simeq Z$, so that the electronic
fraction $Y_{e}$ will be close to half. We also consider the following
parameters values which are: first, when the \textit{SNP} solution is the 
\textit{LMA} scenario, the present parameters are $\sin ^{2}2\theta _{\odot
}=\sin ^{2}2\theta _{12}=0.91$ and $\Delta m_{21}^{2}=6.9\times 10^{-5}\unit{%
eV}^{2}$, while for the \textit{LOW} scenario, we take $\sin ^{2}2\theta
_{12}=0.92$ and $\Delta m_{21}^{2}=1.3\times 10^{-7}\unit{eV}^{2}$.
Secondly, for the atmospheric data, we have $\sin ^{2}2\theta _{atm}=\sin
^{2}2\theta _{23}=1$ and $\Delta m_{32}^{2}=2.7\times 10^{-3}\unit{eV}^{2}$.
Thirdly, for the \textit{1-3} mixing, we choose a value just below the
experimental limit, which is given by \textit{CHOOZ} \cite{Chooz}, and
satisfies the adiabaticity condition at the \textit{H} resonance, which
leads us to take $\sin ^{2}\theta _{13}=10^{-3}$.

{\Large Neutrino Resonances}

The transition between possible neutrino states occur at some preferred
regions, called resonance regions, which are characterized by the equality
between two Hamiltonian's diagonal elements. Thus the possible transitions
are of two kinds: the \textit{MSW} kind occurring due to changing electronic
density when both diagonal elements corresponds to two neutrino states or
two anti-neutrino ones, and the \textit{RSFP} kind corresponding to resonant
spin flavor in presence of a magnetic field \cite{RSFP} and involving a
neutrino and a anti-neutrino state of different flavors. In general we
expect 4 significant resonances, two of each kind namely: the known two 
\textit{MSW} ones which are called \textit{L} and \textit{H} transitions,
and the corresponding \textit{RSFP} ones which we call \textit{\={L}} and 
\textit{\={H}} transitions; in addition to a fifth one of \textit{RSFP}
kind, which we call \textit{A}, but which doesn't affect the neutrino
dynamics. Let us see how we can

deduce these four resonances from general considerations on the full matrix.
Starting from the 6$\times $6 matrix with its six diagonal elements to be
equated two by two, we have \textit{C}$_{\text{\textit{6}}}^{\text{\textit{2}%
}}$\textit{=15} combinations. Out of these 15 combinations, 3 can be
eliminated due to the Majorana character which doesn't allow for the
like-flavor spin flip, 2 can be further eliminated due to the smallness of
the $\nu _{\mu }$-$\nu _{\tau }$\ potential difference and likewise for $%
\bar{\nu}_{\mu }$-$\bar{\nu}_{\tau }.$ For the remaining 10 cases, they
constitute five pairs of conjugate transitions, for which there could only
occur a transition from each pair at a time; since the resonance density
computed by equating diagonal elements can be of both signs, while it's
conjugate transition will be necessarily have the inverse sign, and since
physically only the case with positive value is relevant. This leaves us
with five transitions, the fifth one being as mentioned previously not
significant. We will study each of the \textit{MSW} and \textit{RSFP} kinds
separately.

{\large \textit{MSW}\ Resonances}

In general, \textit{MSW} transition between two neutrinos (or
anti-neutrinos) is studied around its resonance and the general form of the 2%
$\times $2 submatrix is: 
\begin{equation}
\frac{1}{2E}\left( 
\begin{array}{cc}
c & b \\ 
b & Y_{e}\times z%
\end{array}
\right)
\end{equation}
where $z=2\alpha \rho E$,\ and the resonance corresponds to $z_{res}=\frac{c%
}{Y_{e}}.$ The transition will be adiabatic, \textit{i.e.} there is no
jumping between eigenstates, for $\gamma _{res}>1$, where $\gamma _{res}$\
is the adiabaticity parameter \cite{Smirnov} at the resonance given in this
case by: 
\begin{equation}
{\small \gamma }_{res}{\small =}\left( \tfrac{E_{na}}{E}\right) ^{\frac{2}{3}%
}  \label{Ena}
\end{equation}
with 
\begin{equation}
E_{na}=\text{$2$}.\allowbreak 74\times 10^{9}\times Y_{{\small e}}^{\frac{%
{\small 1}}{2}}\times \left( \tfrac{\left\vert b\right\vert }{1\unit{eV}^{2}}%
\right) ^{3}\times \left( \tfrac{\left\vert c\right\vert }{1\unit{eV}^{2}}%
\right) ^{-2}\unit{MeV}
\end{equation}
The parameters $b$ and $c$ are given by: 
\[
\begin{tabular}{cccccc}
Mass hiera- & \textit{SNP} sol- & $b$ (\textit{L}) & $c$ (\textit{L}) & $b$ (%
\textit{H}) & $c$ (\textit{H}) \\ 
Normal & $%
\begin{array}{c}
\text{\textit{LMA}} \\ 
\text{\textit{LOW}}%
\end{array}
$ & $%
\begin{array}{c}
\text{\textit{3.29}}\times \text{\textit{10}}^{-5} \\ 
\text{\textit{6.31}}\times \text{\textit{10}}^{-8}%
\end{array}
$ & $%
\begin{array}{c}
\text{\textit{1.81}}\times \text{\textit{10}}^{-5} \\ 
\text{\textit{-2.60}}\times \text{\textit{10}}^{-6}%
\end{array}
$ & $%
\begin{array}{c}
\text{\textit{8.46}}\times \text{\textit{10}}^{-5} \\ 
\text{\textit{8.53}}\times \text{\textit{10}}^{-5}%
\end{array}
$ & $%
\begin{array}{c}
\text{\textit{2.67}}\times \text{\textit{10}}^{-3} \\ 
\text{\textit{2.69}}\times \text{\textit{10}}^{-3}%
\end{array}
$ \\ 
Inverted & $%
\begin{array}{c}
\text{\textit{LMA}} \\ 
\text{\textit{LOW}}%
\end{array}
$ & $%
\begin{array}{c}
\text{\textit{-3.29}}\times \text{\textit{10}}^{-5} \\ 
\text{\textit{-6.13}}\times \text{\textit{10}}^{-8}%
\end{array}
$ & $%
\begin{array}{c}
\text{\textit{2.35}}\times \text{\textit{10}}^{-5} \\ 
\text{\textit{2.80}}\times \text{\textit{10}}^{-6}%
\end{array}
$ & $%
\begin{array}{c}
\text{\textit{-8.61}}\times \text{\textit{10}}^{-5} \\ 
\text{\textit{8.53}}\times \text{\textit{10}}^{-5}%
\end{array}
$ & $%
\begin{array}{c}
\text{\textit{-2.72}}\times \text{\textit{10}}^{-3} \\ 
\text{\textit{-2.69}}\times \text{\textit{10}}^{-3}%
\end{array}
$%
\end{tabular}%
\]
$b$ and $c$ are given in $\unit{eV}^{2}$. The sign (-) for the \textit{c}
value means that the conjugate transition occurs instead of the usual one,
for example in the case of inverted mass hierarchy within \textit{LMA}
solution, \textit{c}(\textit{H})=\textit{-2.7}$\times $\textit{10}$^{-3}%
\unit{eV}^{2}$ means that $\bar{\nu}_{1}\longleftrightarrow \bar{\nu}_{3}$\
occurs instead of $\nu _{1}\longleftrightarrow \nu _{3}$ \footnote{%
Note that the \textit{L} resonance transition occurs in the antineutrino
channel for the case of normal mass hierarchy within the \textit{LOW} scheme
(For more details see \textit{Appendix A}).}.

The values of $E_{na}$ are given in $\unit{MeV}$ at each layer within all
scenarios by: 
\[
\begin{tabular}{cccc}
Mass hierarchy & \textit{SNP} solution & \textit{L} & \textit{H} \\ 
\textit{Normal} & $%
\begin{array}{c}
\text{\textit{LMA}} \\ 
\text{\textit{LOW}}%
\end{array}
$ & $%
\begin{array}{c}
\text{\textit{2.11}}\mathit{\times }\text{\textit{10}}^{5} \\ 
\text{\textit{7.\allowbreak 20}}\mathit{\times }\text{\textit{10}}^{-2}%
\end{array}
$ & $%
\begin{array}{c}
\text{\textit{164.\thinspace 6}} \\ 
\text{\textit{166.\thinspace \allowbreak 2}}%
\end{array}
$ \\ 
\textit{Inverted} & $%
\begin{array}{c}
\text{\textit{LMA}} \\ 
\text{\textit{LOW}}%
\end{array}
$ & $%
\begin{array}{c}
\text{\textit{1.\allowbreak 25}}\mathit{\times }\text{\textit{10}}^{5} \\ 
\text{\textit{5.\allowbreak 69}}\mathit{\times }\text{\textit{10}}^{-2}%
\end{array}
$ & $%
\begin{array}{c}
\text{\textit{167.\thinspace \allowbreak 2}} \\ 
\text{\textit{166.\thinspace \allowbreak 2}}%
\end{array}
$%
\end{tabular}%
\]

{\large \textit{RSFP} Resonances}

\textit{RSFP} transitions occurring between a neutrino state and an
anti-neutrino one is also studied around its resonance and is of the form: 
\begin{equation}
\frac{1}{2E}\left( 
\begin{array}{cc}
c & sz \\ 
sz & dz%
\end{array}%
\right)  \label{Zs}
\end{equation}%
but with $s=\tfrac{\mu B}{\alpha \rho }\simeq $1.9$\times $10$^{-9}\times (%
\frac{\mu B_{\perp o}}{\mu _{B}G})$. Likewise, we find that the adiabaticity
parameter is given as in Eq. (\ref{Ena}), but with 
\begin{equation}
E_{na}=1.\allowbreak 89\times 10^{-17}\times d^{-\frac{{\small 5}}{{\small 2}%
}}\times \left( \tfrac{\left\vert c\right\vert }{1\unit{eV}^{2}}\right)
\times \left( \tfrac{\mu B_{\perp o}}{\mu _{B}G}\right) ^{3}\unit{MeV}
\end{equation}%
where $d=$\textit{1-2Y}$_{\text{\textit{e}}}$ for both \textit{\={L}} and 
\textit{\={H}}, \textit{d=1-Y}$_{\text{\textit{e}}}$ for \textit{A}. Note
that $E_{na}$\ is proportional to $\left( \mu B_{\perp o}\right) ^{3}$,
which shows that the adiabaticity of the \textit{RSFP} resonances depends
strongly on the interaction term $\mu B_{\perp o}$. The values of \textit{c}
is given in $\unit{eV}^{2}$. 
\[
\begin{tabular}{ccccc}
Mass hierarchy & \textit{SNP} solution & $c$ (\textit{\={L}}) & $c$ (\textit{%
\={H}}) & $c$ (\textit{A}) \\ 
Normal & $%
\begin{array}{c}
\text{\textit{LMA}} \\ 
\text{\textit{LOW}}%
\end{array}%
$ & $%
\begin{array}{c}
\text{\textit{1.81}}\times \text{\textit{10}}^{-5} \\ 
\text{\textit{-2.60}}\times \text{\textit{10}}^{-6}%
\end{array}%
$ & $%
\begin{array}{c}
\text{\textit{2.67}}\times \text{\textit{10}}^{-3} \\ 
\text{\textit{2.69}}\times \text{\textit{10}}^{-3}%
\end{array}%
$ & $%
\begin{array}{c}
\text{\textit{2.65}}\times \text{\textit{10}}^{-3} \\ 
\text{\textit{2.71}}\times \text{\textit{10}}^{-3}%
\end{array}%
$ \\ 
Inverted & $%
\begin{array}{c}
\text{\textit{LMA}} \\ 
\text{\textit{LOW}}%
\end{array}%
$ & $%
\begin{array}{c}
\text{\textit{2.35}}\times \text{\textit{10}}^{-5} \\ 
\text{\textit{2.80}}\times \text{\textit{10}}^{-6}%
\end{array}%
$ & $%
\begin{array}{c}
\text{\textit{-2.72}}\times \text{\textit{10}}^{-3} \\ 
\text{\textit{-2.69}}\times \text{\textit{10}}^{-3}%
\end{array}%
$ & $%
\begin{array}{c}
\text{\textit{-2.74}}\times \text{\textit{10}}^{-3} \\ 
\text{\textit{-2.71}}\times \text{\textit{10}}^{-3}%
\end{array}%
$%
\end{tabular}%
\]

We give in the following table, the values of $E_{na}$ in $\unit{MeV}$
multiplied by $(\frac{\mu B_{\perp o}}{\mu _{B}G})^{-3}$ for \textit{\={L} }%
and\textit{\ \={H}, }and by\textit{\ }$(\frac{\mu _{\mu \tau }B_{\perp o}}{%
\mu _{B}G})^{-3}$ for \textit{A}, at each layer within all scenarios: 
\[
\begin{tabular}{ccccc}
Mass hierarchy & \textit{SNP} solution & \textit{\={L}} & \textit{\={H}} & 
\textit{A} \\ 
\textit{Normal} & $%
\begin{array}{c}
\text{\textit{LMA}} \\ 
\text{\textit{LOW}}%
\end{array}
$ & $%
\begin{array}{c}
\text{\textit{3.\thinspace \allowbreak 38}}\mathit{\times }\text{\textit{10}}%
^{-11} \\ 
\allowbreak \text{\textit{4.\thinspace \allowbreak 86}}\mathit{\times }\text{%
\textit{10}}^{-12}%
\end{array}
$ & $%
\begin{array}{c}
\text{\textit{4.\thinspace \allowbreak 99}}\mathit{\times }\text{\textit{10}}%
^{-9} \\ 
\text{\textit{5.\thinspace \allowbreak 02}}\mathit{\times }\text{\textit{10}}%
^{-9}%
\end{array}
$ & $%
\begin{array}{c}
\text{\textit{\allowbreak 2.\thinspace \allowbreak 83}}\mathit{\times }\text{%
\textit{10}}^{-19} \\ 
\text{\textit{2.\thinspace \allowbreak 90}}\mathit{\times }\text{\textit{10}}%
^{-19}%
\end{array}
$ \\ 
\textit{Inverted} & $%
\begin{array}{c}
\text{\textit{LMA}} \\ 
\text{\textit{LOW}}%
\end{array}
$ & $%
\begin{array}{c}
\text{\textit{4.\thinspace \allowbreak 39}}\mathit{\times }\text{\textit{10}}%
^{-11} \\ 
\text{\textit{5.\thinspace \allowbreak 23}}\mathit{\times }\text{\textit{10}}%
^{-12}%
\end{array}
$ & $%
\begin{array}{c}
\text{\textit{5.\thinspace \allowbreak 08}}\mathit{\times }\text{\textit{10}}%
^{-9} \\ 
\text{\textit{5.\thinspace \allowbreak 02}}\mathit{\times }\text{\textit{10}}%
^{-9}%
\end{array}
$ & $%
\begin{array}{c}
\text{\textit{2.\thinspace \allowbreak 93}}\mathit{\times }\text{\textit{10}}%
^{-19} \\ 
\text{\textit{2.\thinspace \allowbreak 90}}\mathit{\times }\text{\textit{10}}%
^{-19}%
\end{array}
$%
\end{tabular}%
\]

Note that stating that the resonance corresponds to the equality of diagonal
elements of Eq. (\ref{H=}) is not correct strictly speaking. The resonances,
which corresponds to the minima of the difference between the Hamiltonian
eigenvalues, match up exactly the diagonal elements when the off-diagonal
elements are $r-$independent, which is the case for the \textit{MSW} effect.
On the other hand, the resonances will be shifted when the off-diagonal
elements are $r$-dependent as in the case of the \textit{RSFP} effect. In
our case however where the \textit{r}-dependency is of the form $\rho
\propto B\sim r^{-3}$, this shift is negligeable as we show in \textit{%
Appendix B}.

{\Large The Adiabaticity at Neutrino Resonances}

At the resonance, jumping probability is given by \cite{Landau} 
\begin{equation}
P_{f}=exp\left\{ {\small -}\tfrac{\pi }{{\small 2}}\left[ \tfrac{{\small E}%
_{na}}{{\small E}}\right] ^{\frac{2}{3}}\right\}
\end{equation}%
which depends on the energy $E$ as we see from Fig. 2.

\begin{center}
\bigskip

\FRAME{ftbpFU}{9.1028cm}{5.6409cm}{0pt}{\Qcb{The dependence of the flip
probability on the neutrino energy.}}{}{fig2.jpg}{\special{language
"Scientific Word";type "GRAPHIC";maintain-aspect-ratio TRUE;display
"USEDEF";valid_file "F";width 9.1028cm;height 5.6409cm;depth
0pt;original-width 8.6948in;original-height 5.361in;cropleft "0";croptop
"0.9996";cropright "0.9996";cropbottom "0";filename
'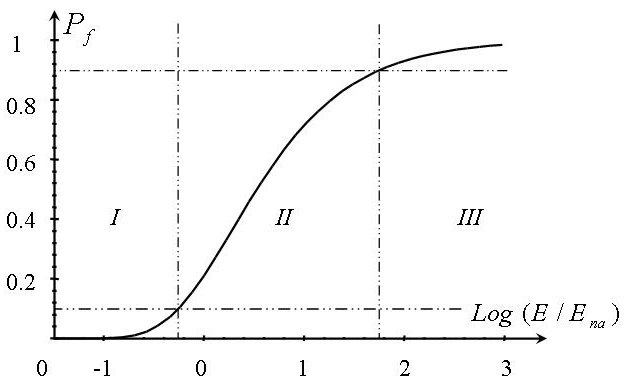';file-properties "XNPEU";}}
\end{center}

Now the observable part of the supernova neutrino spectrum lies mainly
between the energies \textit{5} and \textit{50} $\unit{MeV}$ and we will
consider only energies in this range in our work. One can then divide the
whole range of energy in three parts:

$\bullet $ Part I: $P_{f}\leq 0.1\sim 0$, corresponding to $E_{na}\gtrsim 70%
\unit{MeV}$, and where\ pure adiabatic conversions occurs.

$\bullet $ Part II: $70\unit{MeV}\gtrsim E_{na}\gtrsim 0.2\unit{MeV}$, for
which $0.1\leq P_{f}\leq 0.9.$ In this range, $P_{f}$ increases strongly
with the neutrino energy. The adiabaticity is partially broken.

$\bullet $ Part III: $P_{f}\geq 0.9\sim 1$, corresponding to $E_{na}\lesssim
0.2\unit{MeV}$, where the flip probability is close to $1$, and which leads
to a strong violation of adiabaticity.

Then one can find easily that the transition at the \textit{H} resonance is
completely adiabatic for all possible scenarios, as mentioned above. While,
at the \textit{L} resonance, the transition is: completely adiabatic within 
\textit{LMA} scenario, and therefore $P_{L}\simeq 0$; and completely
non-adiabatic within \textit{LOW} one and therefore $P_{L}\simeq 1$, for
both normal and inverted hierarchies.

For \textit{RSFP} transitions, the adiabaticity depends on the value of $\mu
B_{\perp o}$, and therefore we will divide the whole range of $\mu B_{\perp
o}$\ according to the adiabaticity at the \textit{\={H}} resonance, then the 
$\mu B_{\perp o}$\ regions are:

$I-$ Completely non-adiabatic if: $\mu B_{\perp o}<340\mu _{B}G$.

$II-$ Partially broken if: $340\mu _{B}G<\mu B_{\perp o}<2400\mu _{B}G$.

$III-$ Purely adiabatic if: $\mu B_{\perp o}>2400\mu _{B}G$.

We have plotted in Fig. 3 the flip probability at the \textit{\={H}} layer
as a function of the energy for different values of $\mu B_{\perp o}$. The
curves from (1) to (6) correspond to $\frac{\mu B_{\bot o}}{\mu _{B}G}$=300,
500, 1000, 1500, 2000, 2500 respectvely. 

\begin{center}
$\FRAME{itbpFU}{10.6097cm}{7.5037cm}{0cm}{\Qcb{{\protect\small The
dependance of the flip probability at the layer }$\overline{H}$%
{\protect\small , on the neutrino energy for different values of }$\protect%
\mu B_{\bot o}${\protect\small .}}}{}{fig3.jpg}{\special{language
"Scientific Word";type "GRAPHIC";maintain-aspect-ratio TRUE;display
"USEDEF";valid_file "F";width 10.6097cm;height 7.5037cm;depth
0cm;original-width 9.1385in;original-height 6.4446in;cropleft "0";croptop
"1";cropright "1";cropbottom "0";filename
'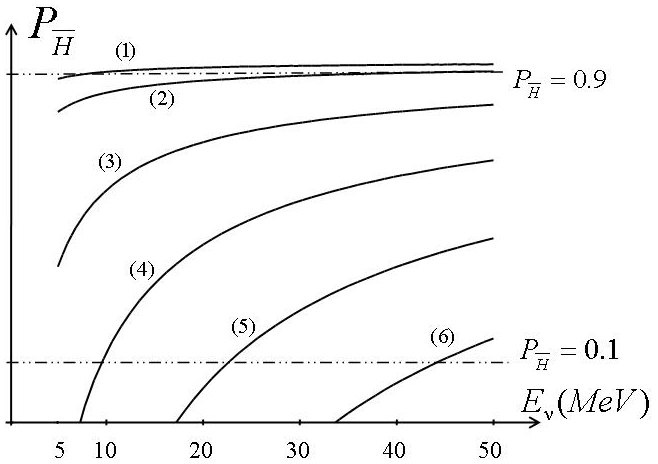';file-properties "XNPEU";}}$
\end{center}

At the \textit{A} resonance layer, the $E_{na}$ values are very small within
the $\mu B_{\perp o}$ considered range, which corresponds to a complete
non-adiabatic conversion, from which we deduce that the transition at the
layer \textit{A} doesn't affect the neutrino flavor dynamics. For the 
\textit{\={L}} layer, the adiabaticity depends on the \textit{SNP} solution
and the specific mass hierarchy. Thus, if $\mu B_{\perp o}$\ is larger than
a given value ($\mu B_{\perp o}$)$_{2}$, the transition will be purely
adiabatic; if it is less than the other limiting value ($\mu B_{\perp o}$)$%
_{1}$, the transition will be completely non-adiabatic; and finally when $%
\mu B_{\perp o}$\ lies between these two values, the adiabaticity will be
partially broken. We give in units of $\mu _{B}G$, these two limiting values
for the various scenarios: 
\[
\begin{tabular}{cccc}
Mass Hierarchy & SNP\ solution & ($\mu B_{\perp o}$)$_{1}$ & ($\mu B_{\perp
o}$)$_{2}$ \\ 
\textit{Normal} & $%
\begin{array}{c}
\text{\textit{LMA}} \\ 
\text{\textit{LOW}}%
\end{array}%
$ & $%
\begin{array}{c}
\text{\textit{1800}} \\ 
\text{\allowbreak \textit{3500}}%
\end{array}%
$ & $%
\begin{array}{c}
\text{\textit{12700}} \\ 
\text{\allowbreak \textit{24300}}%
\end{array}%
$ \\ 
\textit{Inverted} & $%
\begin{array}{c}
\text{\textit{LMA}} \\ 
\text{\textit{LOW}}%
\end{array}%
$ & $%
\begin{array}{c}
\text{\textit{1700}} \\ 
\text{\textit{3400}}%
\end{array}%
$ & $%
\begin{array}{c}
\text{\textit{11600}} \\ 
\text{\textit{23700}}%
\end{array}%
$%
\end{tabular}%
\]%
Note that the $\mu B_{\perp o}$\ interval where the \textit{\={L}} (\textit{%
\={H}}) adiabaticity is partially broken\ intersects with the similar one
for the \textit{\={H}} (\textit{\={L}}) resonance within the \textit{LMA}
solution, but it doesn't within the \textit{LOW} scheme. Then the region 
\textit{II} can be divided in two parts\footnote{%
We will denote them as \textit{II-a} and \textit{II-b}.}: the first one
corresponds to the completely non-adiabatic resonance at the \textit{\={L}}
layer, \textit{i.e.} $\mu B_{\perp o}$ lies between \textit{340}$\mu _{B}$%
\textit{G} and \textit{1800}$\mu _{B}$\textit{G} (\textit{1700}$\mu _{B}$%
\textit{G} for the case inverted hierarchy); while for the remaining part
from \textit{1800}$\mu _{B}$\textit{G} (\textit{1700}$\mu _{B}$\textit{G})
to \textit{2400}$\mu _{B}$\textit{G}, it corresponds to the partially broken
adiabaticity at the layer \textit{\={L}}.

For this purpose, we will give the level crossing for each one of different
scenario, which are given in Fig's 4, 5, 6 and 7, We will check later where
the transition occur, namely whether is it in the neutrino or anti-neutrino
channel (for \textit{MSW}), and which one of the two \textit{RSFP}
conjugated transitions is involved.

The hierarchy of the densities of the resonance layers (\textit{L} with 
\textit{H,} and therefore \textit{\={L}} with \textit{\={H}} too), leads to
the factorization of the neutrino flavor dynamics, the transitions between
two of resonance layers can be considered independently, and therefore each
transition is reduced to a two neutrino problem \cite{DigheSmirnov}.
Furthermore, the \textit{MSW}-\textit{RSFP} resonances regions
non-overlapping condition, for \textit{L-\={L}}\ or \textit{H-\={H}}
transitions, is given by \cite{Akhmedov} 
\begin{equation}
\text{\textit{L}}_{\rho }\left\vert \tan 2\theta _{\alpha \beta }\right\vert
+\left\vert \frac{2\mu _{\alpha \beta }B_{\bot }\left( r_{1}\right) }{\left( 
\frac{\Delta m^{2}}{2E}\right) }\right\vert \text{\textit{L}}_{\rho
}\lesssim \left\vert r_{2}-r_{1}\right\vert 
\end{equation}%
where $r_{1}$ is the radial position of the \textit{\={L}} (\textit{\={H}})
resonance; and $r_{2}$ is that the \textit{L} (\textit{H}) one, and $L_{\rho
}\equiv \left\vert \frac{d\left( \ln \rho \right) }{dr}\right\vert ^{-1}$, $%
\theta _{\alpha \beta }$\ is the mixing angle between the two flavors $%
\alpha $\ and $\beta $ ($\alpha $\textit{,}$\beta $\textit{=e,}$\mu $\textit{%
,}$\tau $). This condition is satisfied for both \textit{L}-\textit{\={L}}\
and \textit{H}-\textit{\={H}}.\FRAME{ftbpFU}{12.1144cm}{8.1736cm}{0pt}{\Qcb{%
The level crossing diagram for the case of normal mass hierarchy within LMA
scheme.}}{}{fig4a.jpg}{\special{language "Scientific Word";type
"GRAPHIC";maintain-aspect-ratio TRUE;display "USEDEF";valid_file "F";width
12.1144cm;height 8.1736cm;depth 0pt;original-width 9.3331in;original-height
6.2777in;cropleft "0";croptop "1";cropright "1";cropbottom "0";filename
'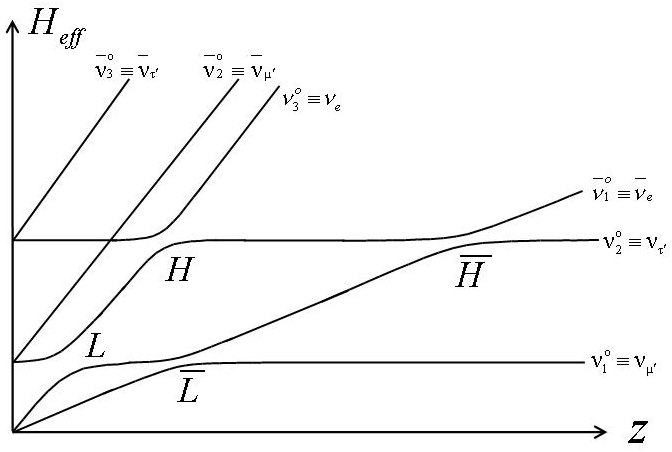';file-properties "XNPEU";}}

\begin{center}
\FRAME{ftbpFU}{12.1166cm}{8.2901cm}{0pt}{\Qcb{The level crossing diagram for
the case of normal mass hierarchy within LOW scheme.}}{}{fig4b.jpg}{\special%
{language "Scientific Word";type "GRAPHIC";maintain-aspect-ratio
TRUE;display "USEDEF";valid_file "F";width 12.1166cm;height 8.2901cm;depth
0pt;original-width 9.4446in;original-height 6.4446in;cropleft "0";croptop
"1";cropright "1";cropbottom "0";filename
'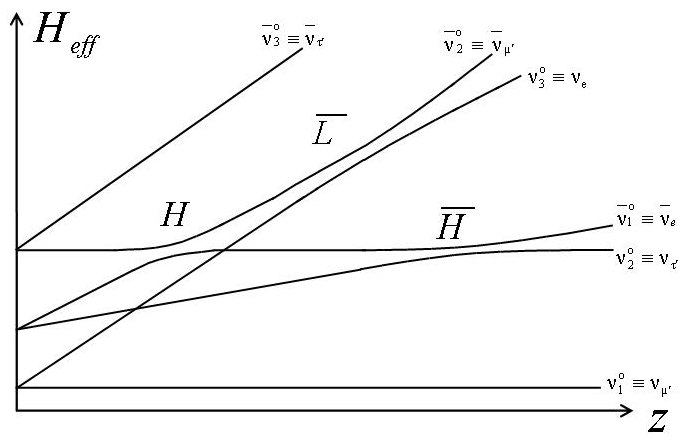';file-properties "XNPEU";}}

\FRAME{ftbpFU}{12.1144cm}{9.105cm}{0pt}{\Qcb{The level crossing diagram for
the case of inverted mass hierarchy within LMA scheme.}}{}{fig5a.jpg}{%
\special{language "Scientific Word";type "GRAPHIC";maintain-aspect-ratio
TRUE;display "USEDEF";valid_file "F";width 12.1144cm;height 9.105cm;depth
0pt;original-width 9.9998in;original-height 7.4996in;cropleft "0";croptop
"1";cropright "1";cropbottom "0";filename
'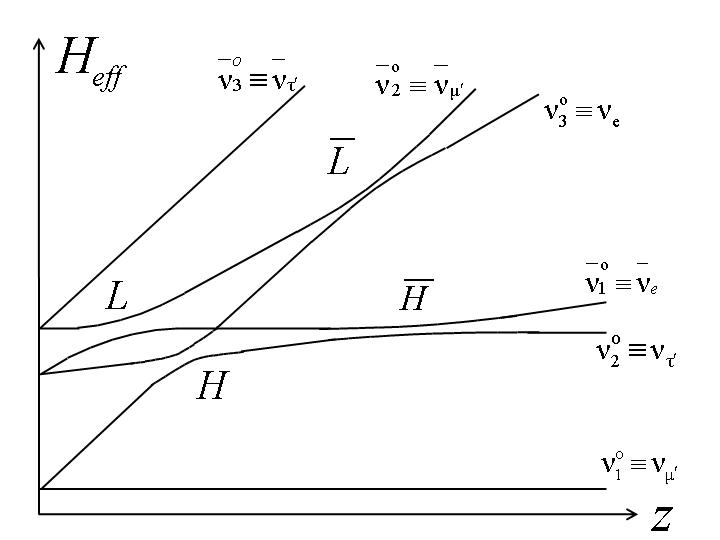';file-properties "XNPEU";}}

\FRAME{ftbpFU}{12.1166cm}{7.8419cm}{0pt}{\Qcb{The level crossing diagram for
the case of inverted mass hierarchy within LOW scheme.}}{}{fig5b.jpg}{%
\special{language "Scientific Word";type "GRAPHIC";maintain-aspect-ratio
TRUE;display "USEDEF";valid_file "F";width 12.1166cm;height 7.8419cm;depth
0pt;original-width 9.5553in;original-height 6.167in;cropleft "0";croptop
"1";cropright "1";cropbottom "0";filename
'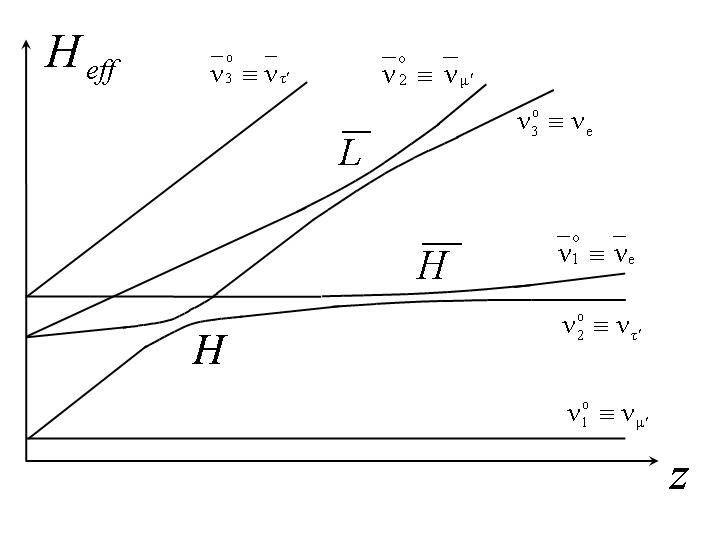';file-properties "XNPEU";}}
\end{center}

{\Large Neutrino Fluxes from Supernovae}

Having constructed the level crossing for the various scenarios, we can now
extract the final $\nu $-flux in terms of $F_{i}$ (or $F_{\overline{i}}$) in
function of the original flux. We can then rewrite it in terms of the
flavored $\nu $-flux using the identity: 
\begin{equation}
F_{l}=\tsum\nolimits_{i}\left\vert U_{li}\right\vert ^{2}F_{i}\text{ and }F_{%
\bar{l}}=\tsum\nolimits_{i}\left\vert U_{li}\right\vert ^{2}F_{\bar{\imath}}
\label{Fl}
\end{equation}%
where \textit{l=e,}$\mu $\textit{,}$\tau $\textit{;}\ \textit{i=1,2,3;} and $%
U_{li}$ are the mixing matrix elements. The final flux is thus given by the
following general relation: 
\begin{equation}
\begin{array}{c}
F_{i}=\tsum_{l}a_{il}F_{l}^{o}+a_{i\bar{l}}F_{\bar{l}}^{o} \\ 
F_{\overline{i}}=\tsum_{l}a_{\bar{\imath}l}F_{l}^{o}+a_{\bar{\imath}\bar{l}%
}F_{\bar{l}}^{o}%
\end{array}
\label{Fi}
\end{equation}%
where $a_{il}$\ is the probability that a $\nu _{l}$, which is produced
inside \textit{SN} core,\ leaves the supernova as a vacuum eigenstate $\nu
_{i}$. Inserting Eq. (\ref{Fi}) in Eq. (\ref{Fl}), we find for the flavored
neutrino flux: 
\begin{equation}
\left( 
\begin{array}{c}
F_{e} \\ 
F_{\bar{e}}^{o} \\ 
4F_{x}%
\end{array}%
\right) {\small =}\left( 
\begin{array}{ccc}
p_{ee} & p_{e\bar{e}} & 1-p_{ee}-p_{e\bar{e}} \\ 
p_{\bar{e}e} & p_{\bar{e}\bar{e}} & 1-p_{\bar{e}e}-p_{\bar{e}\bar{e}} \\ 
1-p_{ee}-p_{\bar{e}e} & 1-p_{e\bar{e}}-p_{\bar{e}\bar{e}} & 2+p_{ee}+p_{e%
\bar{e}}+p_{\bar{e}e}+p_{\bar{e}\bar{e}}%
\end{array}%
\right) \left( 
\begin{array}{c}
F_{e}^{o} \\ 
F_{\bar{e}}^{o} \\ 
F_{x}^{o}%
\end{array}%
\right)  \label{Fmatrix}
\end{equation}%
where $p_{ee}$ is the survival probability of $\nu _{e}$, \textit{i.e,} the
probability that a $\nu _{e}$\ doesn't change during the core collapse; $p_{%
\bar{e}e}$ the probability that a $\nu _{e}$\ leaves \textit{SN} as a $\bar{%
\nu}_{e}$;\ the index ($^{o}$) refers to the original fluxes. It is those
various matrix elements which characterize the probability of \textit{SN}
neutrino conversion. Those probabilities satisfy the condition: 
\begin{equation}
\tsum\nolimits_{\alpha }p_{\alpha \beta }=\tsum\nolimits_{\beta }p_{\alpha
\beta }=1
\end{equation}%
which are functions of the probabilities $a_{il}$s and $U_{ei}$s, and $%
\alpha $,$\beta =e$, $\overline{e}$, $x$.

{\large Case of Normal Mass Hierarchy}

\textbf{For the }\textit{LMA}\textbf{\ scheme:\ }We find from the level
crossing for the case of the normal mass hierarchy (Fig. 4), that the final
flux is given by

\begin{equation}
\begin{array}{c}
\text{\ }F_{1}=(1-p_{\bar{L}})p_{\bar{H}}F_{\bar{e}}^{o}+(1-(1-p_{\bar{L}%
})p_{\bar{H}})F_{x}^{o},F_{2}=(1-p_{\bar{H}})F_{\bar{e}}^{o}+p_{\bar{H}%
}F_{x}^{o} \\ 
F_{3}=F_{e}^{o},\text{ }F_{\bar{1}}=p_{\bar{H}}p_{\bar{L}}F_{\bar{e}%
}^{o}+(1-p_{\bar{H}}p_{\bar{L}})F_{x}^{o}\text{, }F_{\bar{2}}=F_{x}^{o}\text{%
, }F_{\bar{3}}=F_{x}^{o}%
\end{array}
\label{1}
\end{equation}
then the \textit{SN} $\nu $-conversion probabilities are given by 
\begin{equation}
\begin{array}{c}
p_{ee}\simeq 10^{-3},\text{ }p_{e\bar{e}}\simeq 0.65\times \left( 1-p_{\bar{L%
}}\right) p_{\bar{H}}+0.35\left( 1-p_{\bar{H}}\right) \\ 
p_{\bar{e}e}\simeq 0,\text{ }p_{\bar{e}\bar{e}}\simeq 0.\allowbreak 65\times
p_{\bar{L}}p_{\bar{H}}%
\end{array}%
\end{equation}

\textbf{For the }\textit{LOW}\textbf{\ scheme: }According to Fig. 5, the
final flux with the \textit{LOW} scheme is given by 
\begin{equation}
\begin{array}{l}
F_{1}=p_{\bar{H}}F_{x}^{o}+\left( 1-p_{\bar{H}}\right) F_{\bar{e}}^{o},\text{
}F_{2}=F_{x}^{o},\text{ }F_{3}=p_{\bar{L}}F_{e}^{o}+\left( 1-p_{\bar{L}%
}\right) F_{x}^{o} \\ 
F_{\bar{1}}=p_{\bar{H}}F_{\bar{e}}^{o}+\left( 1-p_{\bar{H}}\right) F_{x}^{o},%
\text{ }F_{\bar{2}}=\left( 1-p_{\bar{L}}\right) F_{e}^{o}+p_{\bar{L}%
}F_{x}^{o},\text{ }F_{\bar{3}}=F_{x}^{o}%
\end{array}
\label{2}
\end{equation}%
and therefore the \textit{SN} $\nu $-conversion probabilities are 
\begin{equation}
\begin{array}{c}
p_{ee}\simeq 10^{-3}\times p_{\bar{L}},\text{ }p_{e\bar{e}}\simeq
0.\,\allowbreak 64\times \left( 1-p_{\bar{H}}\right) \\ 
p_{\bar{e}e}\simeq 0.36\times \left( 1-p_{\bar{L}}\right) ,\text{ }p_{\bar{e}%
\bar{e}}\simeq 0.64\times p_{\bar{H}}%
\end{array}%
\end{equation}%
$\allowbreak \allowbreak $

{\large Case of Inverted Mass Hierarchy}

\textbf{For the }\textit{LMA}\textbf{\ scheme: }The final flux is given in
this case (See Fig. 6) by: 
\begin{equation}
\begin{array}{c}
F_{1}=p_{\bar{H}}F_{x}^{o}+\left( 1-p_{\bar{H}}\right) F_{\bar{e}}^{o},\text{
}F_{2}=F_{x}^{o},\text{ }F_{3}=p_{\bar{L}}F_{e}^{o}+\left( 1-p_{\bar{L}%
}\right) F_{x}^{o} \\ 
F_{\bar{1}}=p_{\bar{L}}F_{x}^{o}+\left( 1-p_{\bar{L}}\right) F_{e}^{o},\text{
}F_{\bar{2}}=\left( 1-p_{\bar{H}}\right) F_{x}^{o}+p_{\bar{H}}F_{\bar{e}%
}^{o},\text{ }F_{\bar{3}}=F_{x}^{o}%
\end{array}
\label{3}
\end{equation}%
the \textit{SN} $\nu $-conversion probabilities are given by 
\begin{equation}
\begin{array}{c}
p_{ee}\simeq 10^{-3}\times p_{\bar{L}},\text{ }p_{e\bar{e}}\simeq 0.65\times
\left( 1-p_{\bar{H}}\right) \\ 
p_{\bar{e}e}\simeq 0.65\times \left( 1-p_{\bar{L}}\right) ,\text{ }p_{\bar{e}%
\bar{e}}\simeq 0.35\times p_{\bar{H}}%
\end{array}%
\end{equation}

\textbf{For the }\textit{LOW}\textbf{\ scheme: }We find for the final flux
from Fig. 7: 
\begin{equation}
\begin{array}{l}
F_{1}=p_{\bar{L}}F_{e}^{o}+\left( 1-p_{\bar{L}}\right) F_{x}^{o},\text{ }%
F_{2}=F_{x}^{o},\text{ }F_{3}=\left( 1-p_{\bar{H}}\right) F_{\bar{e}}^{o}+p_{%
\bar{H}}F_{x}^{o} \\ 
F_{\bar{1}}=p_{\bar{L}}F_{x}^{o}+\left( 1-p_{\bar{L}}\right) F_{e}^{o},\,%
\text{\ }F_{\bar{2}}=\left( 1-p_{\bar{H}}\right) F_{x}^{o}+p_{\bar{H}}F_{%
\bar{e}}^{o},\text{ }F_{\bar{3}}=F_{x}^{o}%
\end{array}
\label{4}
\end{equation}%
and the \textit{SN} $\nu $-conversion probabilities are given by: 
\begin{equation}
\begin{array}{c}
p_{ee}\simeq 0.64\times p_{\bar{L}},\text{ }p_{e\bar{e}}\simeq 10^{-3}\times
\left( 1-p_{\bar{H}}\right) \\ 
p_{\bar{e}e}\simeq 0.64\times \left( 1-p_{\bar{L}}\right) ,\text{ }p_{\bar{e}%
\bar{e}}\simeq 0.\allowbreak 36\times p_{\bar{H}}%
\end{array}%
\end{equation}%
Note that any significant value for $p_{e\bar{e}}$\ or $p_{\bar{e}e}$, is a
strong signature of the spin flavor precession effect on the \textit{SN}
neutrino dynamics.

\section{Neutrino Conversion Effects on the Mass Spectrum}

Let us discuss the signatures of the various scenarios as far as the
emerging neutrino signal, as well as their discriminative power at the
future neutrino detectors.

For future supernova neutrino burst, the present detectors can gives, for a
typical supernova, at Super-Kamiokande \cite{SK}\ about \textit{5000} $\nu
_{e}$ events, and a few hundred events can be detected in both SNO \cite{SNO}%
, LVD \cite{LVD} and MACRO \cite{Macro}. In this paper, we are interested on
the features of the final neutrino spectra that are required for the
identification of the neutrino mass spectra. The effects of neutrino mixing
and magnetic moment on the final neutrino spectra can be observed through: 
\textit{a)} the partial or complete disappearance/appearance of the $\nu
_{e} $ neutronization peak, \textit{b)} the appearance of soft, hard or
composite spectra of $\nu _{e}$ and $\bar{\nu}_{e}$; and \textit{c)} the
Earth matter effects on both $\nu _{e}$ and $\bar{\nu}_{e}$ spectra. Let us
estimate these effects on the observed spectra at the Earth detectors.

\textit{a)} \textbf{Neutronization} \textbf{Peak:} It comes at the first
stage of core collapse and corresponds to a and the observed signal during
the of the neutrino burst first few milliseconds duration. In the absence of
neutrino conversion, the dominant signal are $\nu _{e}$s, are produced by
the electron capture on protons and nuclei while the shock wave passes
through the neutrinosphere \cite{Suzuki}. Since the original flux is made of 
$\nu _{e}$, the final observed fluxes give a direct measurement of the
extent of conversion of $\nu _{e}$ into the other neutrino species. It is
thus clear that the neutronization peak is proportional to the $\nu _{e}$s
survival\ probability.

\textit{b)}\textbf{\ The Nature of the Final Spectra:} due to the difference
between the interactions strengths of the various neutrino species with
matter, their average energies differ and are given by \cite{Raffelt}: 
\begin{equation}
\left\langle E_{\nu _{e}}^{o}\right\rangle =10\sim 12\unit{MeV},\text{ }%
\left\langle E_{\bar{\nu}_{e}}^{o}\right\rangle =14\sim 17\unit{MeV},\text{ }%
\left\langle E_{\nu _{x}}^{o}\right\rangle =24\sim 27\unit{MeV}
\end{equation}
E means the original average energy (\textit{i.e.} in the absence $\nu $%
-conversion). Thus finding for example $\left\langle E_{\nu
_{e}}\right\rangle >\left\langle E_{\bar{\nu}_{e}}\right\rangle $,would be
is a signature of neutrino conversion, since it implies that the
contribution of the converted original hard $\nu _{x}$ spectrum to the final 
$\nu _{e}$ flux is significantly larger than its contribution to the final $%
\bar{\nu}_{e}$ flux. On the other hand there is another effect which has to
be accounted for: the fact that the neutrino interaction cross section
increases with energy, the neutrino spectra from the cooling stage won't be
exactly thermal, but will get pinched. We can account easily for this
pinching effect by parametrizing the original spectra as \cite%
{Raffelt,Janka1}: 
\begin{equation}
F_{i}^{o}(E)\propto \frac{E^{2}}{1+\exp (E/T_{i}-\eta _{i})}
\end{equation}
where $T_{i}$\ and $\eta _{i}$\ are given by 
\begin{equation}
\begin{array}{c}
T_{e}\approx 3\sim 4\unit{MeV}\text{, }T_{\overline{e}}\approx 5\sim 6\unit{%
MeV}\text{, }T_{x}\approx 7\sim 9\unit{MeV} \\ 
\eta _{e}\approx 3\sim 5\text{, }\eta _{\overline{e}}\approx 2\sim 2.5\text{%
, }\eta _{x}\approx 0\sim 2%
\end{array}%
\end{equation}

The final $\nu_{e}$ $\left( \bar{\nu}_{e}\right) $ spectrum can be
qualitatively divided into three types:

1) The original \textit{``soft''} spectrum of the $\nu_{e}$ $\left( \bar{\nu 
}_{e}\right) $ (corresponding to the survival probability $p_{ee}=1$ ($P_{%
\bar{e}\bar{e}}=1$)).

2) The \textit{``hard''} spectrum of $\nu _{x}$ (corresponding to the
survival probability $p_{ee}=p_{e\bar{e}}=0$ ($p_{\bar{e}\bar{e}}=p_{\bar{e}%
e}=0$) which would be the case when there is a complete interchange of
spectra, \textit{i.e}. $\nu _{e}\leftrightarrow \nu _{x}$ ($\bar{\nu}%
_{e}\leftrightarrow \nu _{x}$); and not $\bar{\nu}_{e}\leftrightarrow \nu
_{e}$).

$3$) The \textit{``composite''} spectrum, which is a mixture of the original
soft and the hard spectra is comparable in proportions. There are other
cases which are difficult to distinguish like the case where the $\nu _{e}$\
($\bar{\nu}_{e}$) spectrum contains only $\nu _{e}$ and $\nu _{x}$, $\bar{\nu%
}_{e}$\ and $\nu _{x}$, or all species. In order to distinguish among them,
we denote the spectrum containing: $\nu _{e}$ and $\nu _{x}$ by \textit{%
compo-1,} the one containing $\bar{\nu}_{e}$\ and $\nu _{x}$ by \textit{%
compo-2}$;$ and the one with all the species by \textit{compo-3}. In some
cases, the appearance of one of these case is a strong signature of the spin
flavor precession effect, namely when the $\nu _{e}$-spectrum contains $\bar{%
\nu}_{e}$\ and vice-versa.

\textbf{c) the Earth Matter Effect:}

The \textit{SN} neutrinos, in order to reach the detector, have to go
through various amount of the Earth material. This amount depends on the
direction of the \textit{SN} with respect to the Earth as well as the time
of the day. This effect which could modify significantly the flavor
composition of the flux may be sought by comparing the signal of a future 
\textit{SN} from detectors at different geographical locations. Certain
features of the energy spectra could also reveal this effect even from the
observations at one detector.

Neutrinos are expected to be arriving at the surface of the Earth as mass
eigenstates (vacuum eigenstates); where they oscillate in the Earth matter, 
\textit{i.e}, they lose their coherence. The possibility of the Earth matter
effect's observation depends strongly on the differences between the $a_{il}$
and also the differences between the $P_{ie}$ parameters (for more details,
see \textit{Appendix C}).

Let us now summarize the results within all previous scenarios in the
following table:

\begin{center}
\begin{table}[h]
$ 
\begin{tabular}{|c|c|c|c|c|}
\hline
$%
\begin{array}{c}
\text{Region} \\ 
\text{of }\mu B_{\bot o}%
\end{array}
$ & SNP\ solution & $%
\begin{array}{c}
\text{Neutro-} \\ 
\text{Peak}%
\end{array}
$ & $%
\begin{array}{c}
\text{Spectrum of} \\ 
\nu _{e}\text{ \ \ \ \ \ \ \ \ \ \ \ \ \ \ \ \ }\bar{\nu}_{e}%
\end{array}
$ & $%
\begin{array}{c}
\text{Earth effects} \\ 
\nu _{e}\text{ \ \ \ \ \ \ }\bar{\nu}_{e}%
\end{array}
$ \\ \hline
$I$ & 
\begin{tabular}{c}
LMA$\text{ } 
\begin{array}{c}
\text{normal} \\ 
\text{inverted}%
\end{array}
$ \\ 
LOW$\text{ } 
\begin{array}{c}
\text{normal} \\ 
\text{inverted}%
\end{array}
$%
\end{tabular}
& 
\begin{tabular}{c}
$%
\begin{array}[t]{c}
\nu _{x} \\ 
\nu _{x}%
\end{array}
$ \\ 
$%
\begin{array}[t]{c}
\nu _{x} \\ 
\nu _{e},\nu _{x}%
\end{array}
$%
\end{tabular}
& 
\begin{tabular}{c|c}
$%
\begin{array}[t]{c}
\text{hard} \\ 
\text{hard}%
\end{array}
$ & $%
\begin{array}[t]{c}
\text{compo-2} \\ 
\text{compo-2}%
\end{array}
$ \\ 
$%
\begin{array}[t]{c}
\text{hard} \\ 
\text{compo-1}%
\end{array}
$ & $%
\begin{array}[t]{c}
\text{compo-2} \\ 
\text{compo-2}%
\end{array}
$%
\end{tabular}
& 
\begin{tabular}{c}
$%
\begin{array}[t]{c}
\times \text{ \ \ \ \ \ \ }\checkmark \\ 
\times \text{ \ \ \ \ \ \ }\checkmark%
\end{array}
$ \\ 
$%
\begin{array}[t]{c}
\times \text{ \ \ \ \ \ \ }\checkmark \\ 
\checkmark \text{ \ \ \ \ \ \ }\checkmark%
\end{array}
$%
\end{tabular}
\\ \hline
$II$ & $%
\begin{array}{c}
\text{-a LMA } 
\begin{array}{c}
\text{normal} \\ 
\text{inverted}%
\end{array}
\\ 
\text{-b LMA } 
\begin{array}{c}
\text{normal} \\ 
\text{inverted}%
\end{array}
\\ 
\text{LOW } 
\begin{array}{c}
\text{normal} \\ 
\text{inverted}%
\end{array}%
\end{array}
$ & 
\begin{tabular}{c}
$%
\begin{array}[t]{c}
\bar{\nu}_{e},\nu _{x} \\ 
\bar{\nu}_{e},\nu _{x}%
\end{array}
$ \\ 
$%
\begin{array}[t]{c}
\bar{\nu}_{e},\nu _{x} \\ 
\bar{\nu}_{e},\nu _{x}%
\end{array}
$ \\ 
$%
\begin{array}[t]{c}
\bar{\nu}_{e},\nu _{x} \\ 
\nu _{e},\nu _{x}%
\end{array}
$%
\end{tabular}
& 
\begin{tabular}{c|l}
$%
\begin{array}[t]{c}
\text{compo-2} \\ 
\text{compo-2}%
\end{array}
$ & $%
\begin{array}[t]{c}
\text{hard} \\ 
\text{compo-2}%
\end{array}
$ \\ 
$%
\begin{array}[t]{c}
\text{compo-2} \\ 
\text{compo-2}%
\end{array}
$ & $%
\begin{array}[t]{c}
\text{compo-2} \\ 
\text{compo-3}%
\end{array}
$ \\ 
$%
\begin{array}[t]{c}
\text{compo-2} \\ 
\text{compo-1}%
\end{array}
$ & $%
\begin{array}[t]{c}
\text{compo-2} \\ 
\text{compo-2}%
\end{array}
$%
\end{tabular}
& 
\begin{tabular}{c}
$%
\begin{array}[t]{c}
\checkmark \text{ \ \ \ \ \ \ }\checkmark \\ 
\checkmark \text{ \ \ \ \ \ \ }\checkmark%
\end{array}
$ \\ 
$%
\begin{array}[t]{c}
\checkmark \text{ \ \ \ \ \ \ }\checkmark \\ 
\checkmark \text{ \ \ \ \ \ \ }\checkmark%
\end{array}
$ \\ 
$%
\begin{array}[t]{c}
\checkmark \text{ \ \ \ \ \ \ }\checkmark \\ 
\checkmark \text{ \ \ \ \ \ \ }\checkmark%
\end{array}
$%
\end{tabular}
\\ \hline
$III$ & 
\begin{tabular}{c}
LMA$\text{ } 
\begin{array}{c}
\text{normal} \\ 
\text{inverted}%
\end{array}
$ \\ 
LOW$\text{ } 
\begin{array}{c}
\text{normal} \\ 
\text{inverted}%
\end{array}
$%
\end{tabular}
& 
\begin{tabular}{c}
$%
\begin{array}[t]{c}
\bar{\nu}_{e},\nu _{x} \\ 
\nu _{e},\nu _{x}%
\end{array}
$ \\ 
$%
\begin{array}[t]{c}
\bar{\nu}_{e},\nu _{x} \\ 
\nu _{e},\nu _{x}%
\end{array}
$%
\end{tabular}
& 
\begin{tabular}{c|c}
$%
\begin{array}[t]{c}
\text{compo-2} \\ 
\text{compo-2}%
\end{array}
$ & $%
\begin{array}[t]{c}
\text{hard} \\ 
\text{compo-2}%
\end{array}
$ \\ 
$%
\begin{array}[t]{c}
\text{compo-2} \\ 
\text{compo-1}%
\end{array}
$ & $%
\begin{array}[t]{c}
\text{hard} \\ 
\text{compo-1}%
\end{array}
$%
\end{tabular}
& 
\begin{tabular}{c}
$%
\begin{array}[t]{c}
\checkmark \text{ \ \ \ \ \ \ }\times \\ 
\checkmark \text{ \ \ \ \ \ \ }\checkmark%
\end{array}
$ \\ 
$%
\begin{array}[t]{c}
\checkmark \text{ \ \ \ \ \ \ }\checkmark \\ 
\checkmark \text{ \ \ \ \ \ \ }\times%
\end{array}
$%
\end{tabular}
\\ \hline
\end{tabular}
$%
\caption[{\protect\small The dependance of the final spectra on: }1%
{\protect\small ) the }SNP{\protect\small \ solution, }2{\protect\small )
the mass hierarchy and }3{\protect\small ) the range of }$\protect\mu %
B_{\bot o}${\protect\small . \textquotedblleft }$soft${\protect\small %
\textquotedblright\ in the column refers to the original }$\protect\nu _{e}$%
{\protect\small \ (}$\bar{\protect\nu}_{e}${\protect\small ) spectrum and
\textquotedblleft }$hard${\protect\small \textquotedblright\ refers to the
original }$\protect\nu _{x}${\protect\small \ spectrum. In the Earth matter
effect column, }$\checkmark ${\protect\small \ and\ }$\times $%
{\protect\small \ indicate the possibility of significant effects.}]{%
{\protect\small The dependance of the final spectra on: }1{\protect\small )
the }SNP{\protect\small \ solution, }2{\protect\small ) the mass hierarchy
and }3{\protect\small ) the range of }$\protect\mu B_{\bot o}$%
{\protect\small . \textquotedblleft }$soft${\protect\small %
\textquotedblright\ in the column refers to the original }$\protect\nu _{e}$%
{\protect\small \ (}$\bar{\protect\nu}_{e}${\protect\small ) spectrum and
\textquotedblleft }$hard${\protect\small \textquotedblright\ refers to the
original }$\protect\nu _{x}${\protect\small \ spectrum. In the Earth matter
effect column, }$\checkmark ${\protect\small \ and\ }$\times $%
{\protect\small \ indicate the possibility of significant effects.}}
\end{table}
\end{center}

\section{Conclusion}

We attempted in this paper to check the sensibility of the supernovae final
flux of neutrinos to both the magnetic field and matter effect assuming
three active neutrinos. We have taken the value of $\left| U_{e3}\right|
^{2} $\ to be just below the experimental upper bound given by CHOOZ \cite%
{Chooz}, which makes the transition at the \textit{H} layer completely
adiabatic. We then divided the whole range of values considered for $\mu
B_{\bot o}$ into three regions \textit{I, II, }and \textit{III} according to
the adiabaticity at the \textit{\={H}} layer. We could then make some
predictions on the conversion effects for supernova neutrinos. The
predictions differ for the different schemes, which opens up the possibility
of discriminating between them using data from future neutrino bursts. We
studied the possibility of observing the conversion effects through: (%
\textit{i}) The partial or complete change of the flavor of the
neutronization peak, (\textit{ii}) The appearance of a hard or composite $%
\nu _{e}$ and/or $\bar{\nu}_{e}$ spectra due to the conversion effect
instead of their original soft spectra. (\textit{iii}) the Earth matter
effect from the conversion in the Earth material. We found that indeed
neutrino conversion does change significantly the spectrum shape. Let us now
summarize the salient features of the final flux as it appears from the last
table:

1. The appearance of a hard $\bar{\nu}_{e}$ makes the mass hierarchy to be
normal.

2. A hard $\nu _{e}$ spectrum makes $\mu B_{o\bot }$ to be in the first
region, while a hard $\bar{\nu}_{e}$ spectrum exclude the first region.

3. The absence of the Earth matter effect in the $\nu _{e}$\ channel implies
that $\mu B_{o\bot }$ to be in the first region; while the absence in $\bar{%
\nu}_{e}$ one implies that it must be in the third region.

4. If $\bar{\nu}_{e}$ contains all spieces (\textit{i.e.} $\nu _{e}$, $\bar{%
\nu}_{e}$ and $\nu _{x}$), the \textit{SNP} solution must be \textit{LMA},
the mass hierarchy is inverted and $\mu B_{\bot o}$\ lies between \textit{%
1700}$\mu _{B}G$\ and \textit{2400}$\mu _{B}G$.

5. If both spectra of $\nu _{e}$ and $\bar{\nu}_{e}$\ contain only $\nu _{e}$
and $\nu _{x}$, the mass hierarchy will be inverted, the \textit{SNP}
solution will be \textit{LOW}, and $\mu B_{\bot o}$\ will be larger than 
\textit{2400}$\mu _{B}G$.

Notice that some possible observations can rule out all of the various
scenarios above, like the appearance of a soft $\nu _{e}$\ (or $\bar{\nu}%
_{e} $) signal, which would exclude all the above scenarios, since then the
value of $\left\vert U_{e3}\right\vert ^{2}$\ would be less than \textit{10}$%
^{\mathit{-3}}$.

The final neutrino spectra can thus help in resolving three main kinds of
ambiguities that remain to be resolved with the current data: (\textit{i})
the solution of the \textit{SNP}, (\textit{ii}) the type of mass hierarchy
(sign of $\Delta m_{32}^{2}$), and (\textit{iii}) probe the magnitude of the 
$\mu B_{o\bot }$ value, assuming $\left\vert U_{e3}\right\vert ^{2}$ to be 
\textit{10}$^{\mathit{-3}}$. The implications of the results of this work
will depend on when will the next \ neutrino burst from a Galactic supernova
be detected. On the other hand, there is a good chance that within the next
few years the present, ongoing, or future planned experiments will allow us
to identify the specific solution of the solar neutrino problem considered
in this work. This will significantly diminish the number of possible
schemes and will allow us to further sharpen the predictions of the effects
for supernova neutrinos.

\textbf{Acknowledgement}

We would like to thank Francesco Vissani for useful comments on the paper.

\appendix

\section{In which Channel does the \textit{L} Resonance Occur?}

Taking the Hamiltonian in the ($\nu _{e}$,$\nu _{\alpha }$) basis: 
\begin{equation}
\frac{1}{2E}\left( 
\begin{array}{cc}
\text{\textit{W}} & \text{\textit{K}} \\ 
\text{\textit{K}} & \text{\textit{Qz}}%
\end{array}
\right)
\end{equation}
where $\nu _{\alpha }$ is any combination of the non-electronic neutrinos, 
\textit{K} and \textit{W} are given in function of the various $\Delta m^{2}$%
's; and \textit{Q} is a function of $\theta _{ij}$'s and \textit{Y}$_{\text{%
\textit{e}}}$.\ We know that the resonance occurs at \textit{z}$_{\text{%
\textit{res}}}=\frac{W}{Q};$ if $z_{res}<0$,\ then the transition occurs
between $\left( \bar{\nu}_{e},\bar{\nu}_{\alpha }\right) $\ instead of $%
\left( \nu _{e},\nu _{\alpha }\right) $. In general \textit{Q} is always
positive, thus we should check the sign of \textit{W}.

\textit{W} represents the eigenvalues of the Hamiltonian submatrix, which
corresponds to the non-electronic states ($\nu _{\mu }$,$\nu _{\tau }$).
(see. Sec II); In order to probe the sign of these two values in the general
case, we take a $2\times 2$ symmetric matrix, similar to the mass matrix in
the subspace ($\nu _{\mu }$,$\nu _{\tau }$), as follow 
\begin{equation}
\left( 
\begin{array}{cc}
\varepsilon & \lambda \\ 
\lambda & \sigma%
\end{array}
\right)
\end{equation}
where $\varepsilon $, $\lambda $ and $\sigma $ are reels; it's eigenvalues
are: $\tfrac{1}{2}$($\varepsilon +\sigma \pm \sqrt{\left( \varepsilon
-\sigma \right) ^{2}+4\lambda ^{2}}$)$;$ then one can see that the two
eigenvalues have:

$\bullet $ the same sign if $\left| \varepsilon \sigma \right| >\lambda ^{2}$%
.

$\bullet $ different signs if $\left| \varepsilon \sigma \right| <\lambda
^{2}$.

The small value, in absolute value, leads the resonance at the \textit{L}
layer, while the largest one leads to the resonance at the \textit{H} layer.
Thus, we are led to look at the sign of the smallest value, replacing $%
\varepsilon $, $\lambda $ and $\sigma $ by their values in our case, and
neglecting terms multiplied by $\frac{\Delta m_{21}^{2}}{\Delta m_{31}^{2}}$%
, then one can find the condition that the resonances at both \textit{L} and 
\textit{H} layers, occurs in the same channel, it can be written as 
\begin{equation}
\sin ^{2}\theta _{13}\lesssim \tfrac{\Delta m_{21}^{2}}{\Delta m_{32}^{2}}%
\cos 2\theta
\end{equation}
this condition is satisfied in the case of inverted mass hierarchy with both 
\textit{LMA} and \textit{LOW} scenarios. But for the case of normal mass
hierarchy, it is satisfied only within \textit{LMA} scheme.

\section{The Adiabaticity of Neutrino Conversion}

For neutrinos traveling through a non-constant density medium, it is proved
that the adiabaticity is always satisfied except for the cases around the
resonance layer, where it must be studied more thoroughly. The resonance, in
general, can be characterized by the matter density value $\rho $, and
therefore the distance $r$ from the \textit{SN} center, when the difference
between the two Hamiltonian eigenvalues, $\Delta H_{eff}=\left| \tfrac{%
M_{2}^{2}}{2E}-\tfrac{M_{1}^{2}}{2E}\right| $, is minimal, \textit{i.e.} 
\begin{equation}
\tfrac{\partial }{\partial f}\left( \Delta H_{eff}\right| _{res}=0
\end{equation}
where $f$\ is either the density $\rho $ or the travelled distance $r$.
Considering the $2$-$\nu $\ scheme (Like $\nu _{e}$,$\nu _{\mu }$) case for
the \textit{MSW} effect, \textit{H} can be written as: 
\begin{equation}
H=\frac{1}{2E}U\left( 
\begin{array}{cc}
m_{1}^{2} & 0 \\ 
0 & m_{2}^{2}%
\end{array}
\right) U^{\dagger }+\dfrac{\alpha }{2}\rho \left( 
\begin{array}{cc}
3Y_{e}-1 & 0 \\ 
0 & Y_{e}-1%
\end{array}
\right)
\end{equation}
and the mixing matrix \textit{U},\ is given by 
\begin{equation}
U=\left( 
\begin{array}{cc}
\cos \theta & \sin \theta \\ 
-\sin \theta & \cos \theta%
\end{array}
\right)
\end{equation}
Neglecting the terms proportional to the unity matrix, we find 
\begin{equation}
H=U\left( 
\begin{array}{cc}
0 & 0 \\ 
0 & \Delta%
\end{array}
\right) U^{\dagger }+\alpha \rho \left( 
\begin{array}{cc}
1 & 0 \\ 
0 & 0%
\end{array}
\right) =\left( 
\begin{array}{cc}
\alpha \rho +\Delta \sin ^{2}\theta & \tfrac{\Delta }{2}\sin 2\theta \\ 
\tfrac{\Delta }{2}\sin 2\theta & \Delta \cos ^{2}\theta%
\end{array}
\right)
\end{equation}
where $\Delta =\frac{m_{2}^{2}-m_{1}^{2}}{2E}$. Then the difference between
the two eigenvalues is: 
\begin{equation}
\sqrt{\left( \alpha \rho -\Delta \cos 2\theta \right) ^{2}+\Delta ^{2}\sin
^{2}2\theta }
\end{equation}
which takes its minimum at 
\begin{equation}
\rho =\tfrac{\Delta }{\alpha }\cos 2\theta  \label{A6}
\end{equation}
This corresponds to the equality of the diagonal elements.

If we try to redo the same operation but writing this time \textit{H}\ in
the vacuum state basis $\left( \nu _{1},\nu _{2}\right) $, we find 
\begin{equation}
H=U^{\dagger }HU=\left( 
\begin{array}{cc}
0 & 0 \\ 
0 & \Delta%
\end{array}%
\right) +\alpha \rho \left( 
\begin{array}{cc}
\cos ^{2}\theta & \tfrac{1}{2}\sin 2\theta \\ 
\tfrac{1}{2}\sin 2\theta & \sin ^{2}\theta%
\end{array}%
\right)
\end{equation}%
The difference between the eigenvalues takes its minimum at the same value
of $\rho $, given in Eq. (\ref{A6}), because they are different
representations of the same operator \textit{H}, which doesn't correspond to
the diagonal elements equality 
\begin{equation}
\rho =\tfrac{\Delta }{\alpha \cos 2\theta }
\end{equation}

We will obtain the same result if we write \textit{H} in any other basis, 
\textit{i.e.}, the minimum of $\Delta H_{eff}$ doesn't correspond the
diagonal elements equality.

One can deduce that the minimum of $\Delta H_{eff}$\ corresponds to the
diagonal elements equality only when the off-diagonal elements are constant, 
\textit{i.e.} $\rho $-independent which is realized only in the flavored
basis.

There are some cases, like the one in the presence of magnetic moment
interaction, where the flavored basis is not the best basis to deduce the
resonance density directly, \textit{i.e.} by equating the diagonal elements.

Let us consider our case, from Eq. (\ref{Zs}), the minimum of $\Delta
H_{eff} $\ is given by 
\begin{equation}
\tfrac{1}{2E}\sqrt{\left( b-cz\right) ^{2}+4s^{2}z^{2}}
\end{equation}%
which corresponds to 
\begin{equation}
z_{res}=\tfrac{bc}{c^{2}+4s^{2}}
\end{equation}%
but we take it $\frac{b}{c};$ since $s\ll c$, then one can write 
\begin{equation}
z_{res}\simeq \frac{b}{c}\left( 1-7.\,\allowbreak 46\times 10^{-14}\times
\left( \tfrac{\mu B_{\perp o}}{\mu _{B}G}\right) ^{2}\right)
\end{equation}%
thus the layer radius $r_{res}$ is corrected by $\Delta r_{res}$, where $%
\frac{\Delta r_{res}}{r_{res}}=-4.21\times 10^{-5}\times \left( \frac{\mu
B_{\perp o}}{\mu _{B}G}\right) ^{\frac{2}{3}}$, then the adiabaticity
parameter will be corrected by the factor 
\begin{equation}
\xi =1-\allowbreak 1.24\times 10^{-13}\times \left( \tfrac{\mu B_{\perp o}}{%
\mu _{B}G}\right) ^{2}
\end{equation}%
and the jumping probability becomes 
\begin{equation}
P_{f}^{\text{(corrected)}}=\left[ P_{f}\right] ^{\xi }
\end{equation}%
suppose that $P_{f}=$\textit{0.4}, in order to be corrected by \textit{0.1}, 
\textit{i.e.} $P_{f}^{\text{(\textit{corrected})}}=$\textit{0.5}, we need
that the quantity $\mu B_{\perp o}$ to be larger than $1.4\times 10^{6}\mu
_{B}G$; which is a huge value non-available in the supernovae. We conclude
that this correction doesn't affect the neutrino conversion in our work, and
we have thus neglected it.

\section{The Earth Matter Effect\protect\footnote{%
We are here closely following the treatment of \cite{DigheSmirnov}, adopting
it to our case by taking into account the possibility of the $\nu -\bar{\nu}$
transition. This effect is studied\ with more details for difference
scenarios in \cite{LunardiniSmirnov}.}}

If neutrinos reach the Earth detectors without interacting with matter, the
signal at the detectors is given by Eq. (\ref{Fmatrix}). Since they reach
detectors through the Earth matter, the flux given by Eq. (\ref{Fmatrix})
will be modified in general.

Let $P_{ie}$ ($P_{\bar{\imath}\bar{e}}$)\ be the probability that a vacuum
mass eigenstate $\nu _{i}$ ($\bar{\nu}_{i}$)\ entering the Earth reaches the
detector as a $\nu _{e}$ ($\bar{\nu}_{e}$). The flux of $\nu _{e}$ ($\bar{\nu%
}_{e}$) at the detector is 
\begin{equation}
F_{e}^{D}=\tsum\nolimits_{i}P_{ie}F_{i},\text{ \ }F_{\bar{e}%
}^{D}=\tsum\nolimits_{i}P_{\bar{\imath}\bar{e}}F_{\bar{\imath}}
\end{equation}%
Inserting $F_{i}$, we get 
\begin{equation}
\begin{array}{c}
F_{e}^{D}=F_{e}^{o}\tsum\nolimits_{i}P_{ie}a_{ie}+F_{\bar{e}%
}^{o}\tsum\nolimits_{i}P_{ie}a_{i\overline{e}}+F_{x}^{o}\tsum%
\nolimits_{i}P_{ie}\left( 1-a_{ie}-a_{i\bar{e}}\right) \\ 
F_{\bar{e}}^{D}=F_{e}^{o}\tsum\nolimits_{i}P_{\bar{\imath}\bar{e}}a_{\bar{%
\imath}e}+F_{\bar{e}}^{o}\tsum\nolimits_{i}P_{\bar{\imath}\bar{e}}a_{\bar{%
\imath}\bar{e}}+F_{x}^{o}\tsum\nolimits_{i}P_{\bar{\imath}\bar{e}}\left(
1-a_{\bar{\imath}e}-a_{\bar{\imath}\bar{e}}\right)%
\end{array}%
\end{equation}%
where the $a$'s are defined\ in Sec. 3. Then one can write 
\begin{equation}
p_{ee}^{D}=\tsum\nolimits_{i}P_{ie}a_{ie}\text{, \ }p_{e\bar{e}%
}^{D}=\tsum\nolimits_{i}P_{ie}a_{i\bar{e}}\text{, \ }p_{\bar{e}%
e}^{D}=\tsum\nolimits_{i}P_{\bar{\imath}\bar{e}}a_{\bar{\imath}e}\text{,\ }%
p_{\bar{e}\bar{e}}^{D}=\tsum\nolimits_{i}P_{\bar{\imath}\bar{e}}a_{\bar{%
\imath}\bar{e}}
\end{equation}%
Similarly we have 
\begin{equation}
\begin{array}{c}
F_{e}=p_{ee}F_{e}^{o}+p_{e\bar{e}}F_{\bar{e}}^{o}+\left( 1-p_{ee}-p_{e\bar{e}%
}\right) F_{x}^{o} \\ 
F_{\bar{e}}=p_{\bar{e}\bar{e}}F_{\bar{e}}^{o}+p_{\bar{e}e}F_{e}^{o}+\left(
1-p_{\bar{e}\bar{e}}-p_{\bar{e}e}\right) F_{x}^{o}%
\end{array}%
\end{equation}%
which enables us to write: 
\begin{equation}
\begin{array}{c}
F_{e}^{D}=P_{ee}^{D}F_{e}^{o}+P_{e\bar{e}}^{D}F_{\bar{e}}^{o}+\left(
1-P_{ee}^{D}-P_{e\bar{e}}^{D}\right) F_{x}^{o} \\ 
F_{\overline{e}}^{D}=P_{\bar{e}\bar{e}}^{D}F_{\bar{e}}^{o}+P_{\bar{e}%
e}^{D}F_{e}^{o}+\left( 1-P_{\bar{e}\bar{e}}^{D}-P_{\bar{e}e}^{D}\right)
F_{x}^{o}%
\end{array}%
\end{equation}%
From those expressions we deduce: 
\begin{equation}
\begin{array}{c}
F_{e}^{D}-F_{e}^{o}=\left( F_{e}^{o}-F_{x}^{o}\right) \left(
P_{ee}^{D}-p_{ee}\right) +\left( F_{\bar{e}}^{o}-F_{x}^{o}\right) \left( P_{e%
\bar{e}}^{D}-p_{e\bar{e}}\right) \\ 
F_{\bar{e}}^{D}-F_{\bar{e}}^{o}=\left( F_{\bar{e}}^{o}-F_{x}^{o}\right)
\left( P_{\bar{e}\bar{e}}^{D}-p_{\bar{e}\bar{e}}\right) +\left(
F_{e}^{o}-F_{x}^{o}\right) \left( P_{\bar{e}e}^{D}-p_{\bar{e}e}\right)%
\end{array}%
\end{equation}%
so the Earth matter effect can be quantified by the various differences ($%
P_{ee}^{D}-p_{ee}$), ($P_{e\bar{e}}^{D}-p_{e\bar{e}}$), ($P_{\bar{e}\bar{e}%
}^{D}-p_{\bar{e}\bar{e}}$) and ($P_{\bar{e}e}^{D}-p_{\bar{e}e}$), which
equal; 
\begin{equation}
\begin{array}{c}
P_{ee}^{D}-p_{ee}=\tsum a_{ie}(P_{ie}-\left\vert U_{ei}\right\vert ^{2})%
\text{, \ }P_{e\bar{e}}^{D}-p_{e\bar{e}}=\tsum a_{i\bar{e}%
}(P_{ie}-\left\vert U_{ei}\right\vert ^{2}) \\ 
P_{\bar{e}\bar{e}}^{D}-p_{\bar{e}\bar{e}}=\tsum a_{\bar{\imath}\bar{e}}(P_{%
\bar{\imath}\bar{e}}-\left\vert U_{ei}\right\vert ^{2})\text{, }P_{\bar{e}%
e}^{D}-p_{\bar{e}e}=\tsum a_{\bar{\imath}e}(P_{\bar{\imath}\bar{e}%
}-\left\vert U_{ei}\right\vert ^{2})%
\end{array}%
\end{equation}%
One can finally write $P_{ee}^{D}-p_{ee}$ as 
\begin{equation}
P_{ee}^{D}-p_{ee}=a_{1e}(P_{1e}-\left\vert U_{e1}\right\vert
^{2})+a_{2e}(P_{2e}-\left\vert U_{e2}\right\vert
^{2})+a_{3e}(P_{3e}-\left\vert U_{e3}\right\vert ^{2})  \label{c8}
\end{equation}%
We obtain similar expressions for the other difference terms. We notice that
the last term in Eq. (\ref{c8}) is negligeable due to the very small depth
oscillation of $\nu _{3}$\ inside the Earth \cite{DigheSmirnov} 
\begin{equation}
P_{3e}-\left\vert U_{e3}\right\vert ^{2}\leq 10^{-3}
\end{equation}%
Taking into account the relations $\tsum \left\vert U_{ei}\right\vert ^{2}=1$%
\ and $\tsum_{i}P_{ie}=\tsum P_{\bar{\imath}\bar{e}}=1$, one can write 
\begin{equation}
\begin{array}{c}
P_{ee}^{D}-p_{ee}=\left( a_{2e}-a_{1e}\right) (P_{2e}-\left\vert
U_{e2}\right\vert ^{2}),\text{ \ }P_{e\bar{e}}^{D}-p_{e\bar{e}}=\left( a_{2%
\overline{e}}-a_{1\overline{e}}\right) (P_{2e}-\left\vert U_{e2}\right\vert
^{2}) \\ 
P_{\bar{e}\bar{e}}^{D}-p_{\bar{e}\bar{e}}=\left( a_{\bar{2}\bar{e}}-a_{\bar{1%
}\bar{e}}\right) (P_{\bar{2}\bar{e}}-\left\vert U_{e2}\right\vert ^{2}),%
\text{ \ }P_{\bar{e}e}^{D}-p_{\bar{e}e}=\left( a_{\bar{2}e}-a_{\bar{1}%
e}\right) (P_{\bar{2}\bar{e}}-\left\vert U_{e2}\right\vert ^{2})%
\end{array}%
\end{equation}%
finally, the differences $F_{e}^{D}-F_{e}^{o}$\ and $F_{\bar{e}}^{D}-F_{\bar{%
e}}^{o}$\ can be written as 
\begin{equation}
\begin{array}{c}
F_{e}^{D}-F_{e}^{o}\approx \left[ \left( F_{e}^{o}-F_{x}^{o}\right) \left(
a_{2e}-a_{1e}\right) +\left( F_{\bar{e}}^{o}-F_{x}^{o}\right) \left( a_{2%
\bar{e}}-a_{1\bar{e}}\right) \right] (P_{2e}-\left\vert U_{e2}\right\vert
^{2}) \\ 
F_{\bar{e}}^{D}-F_{\bar{e}}^{o}\approx \left[ \left( F_{\bar{e}%
}^{o}-F_{x}^{o}\right) \left( a_{\bar{2}\bar{e}}-a_{\bar{1}\bar{e}}\right)
+\left( F_{e}^{o}-F_{x}^{o}\right) \left( a_{\bar{2}e}-a_{\bar{1}e}\right) %
\right] (P_{\bar{2}\bar{e}}-\left\vert U_{e2}\right\vert ^{2})%
\end{array}%
\end{equation}%
In general, when the signal from two detectors $D_{1}$ and $D_{2}$ are
compared, we get the following flux differences: 
\begin{equation}
\begin{array}{c}
F_{e}^{D_{1}}-F_{e}^{D_{2}}\simeq \left[ \left( F_{e}^{o}-F_{x}^{o}\right)
\left( a_{2e}-a_{1e}\right) +\left( F_{\bar{e}}^{o}-F_{x}^{o}\right) \left(
a_{2\bar{e}}-a_{1\bar{e}}\right) \right] (P_{2e}^{\left( 1\right)
}-P_{2e}^{\left( 2\right) }) \\ 
F_{\bar{e}}^{D_{1}}-F_{\bar{e}}^{D_{2}}\simeq \left[ \left( F_{\bar{e}%
}^{o}-F_{x}^{o}\right) \left( a_{\bar{2}\bar{e}}-a_{\bar{1}\bar{e}}\right)
+\left( F_{e}^{o}-F_{x}^{o}\right) \left( a_{\bar{2}e}-a_{\bar{1}e}\right) %
\right] (P_{\bar{2}\bar{e}}^{\left( 1\right) }-P_{\bar{2}\bar{e}}^{\left(
2\right) })%
\end{array}%
\end{equation}%
According to this relation, the difference $F_{e}^{D_{1}}-F_{e}^{D_{2}}$\ ($%
F_{\bar{e}}^{D_{1}}-F_{\bar{e}}^{D_{2}}$) is factorized: it is proportional
to the difference of the Earth oscillation probability $P_{2e}$ ($P_{\bar{2}%
\bar{e}}$) at the two detectors and a function of supernovae oscillation
probabilities $a$'s and the difference in the original\ fluxes of $\nu _{e}$
($\bar{\nu}_{e}$)\ and $\nu _{x}$.

Let us consider these factors separately:

$\qquad $\textit{1)} $P_{2e}^{\left( 1\right) }-P_{2e}^{\left( 2\right) }$\
and $P_{\bar{2}\bar{e}}^{\left( 1\right) }-P_{\bar{2}\bar{e}}^{\left(
2\right) }$: if the neutrino trajectory go through only the Earth's mantle,
one can use a constant density approximation, which gives \cite{DigheSmirnov}
\begin{equation}
\begin{array}{c}
P_{2e}^{\left( 1\right) }-P_{2e}^{\left( 2\right) }=\sin 2\theta
_{12}^{m}\sin 2(\theta _{12}^{m}-\theta _{12})\left[ \sin ^{2}\left( \tfrac{%
\pi d_{1}}{L_{m}}\right) -\sin ^{2}\left( \tfrac{\pi d_{2}}{L_{m}}\right) %
\right] \\ 
P_{\bar{2}\bar{e}}^{\left( 1\right) }-P_{\bar{2}\bar{e}}^{\left( 2\right)
}=\sin 2\theta _{\bar{e}\bar{2}}^{m}\sin 2(\theta _{\bar{e}\bar{2}%
}^{m}-\theta _{e2})\left[ \sin ^{2}\left( \tfrac{\pi d_{1}}{L_{m}}\right)
-\sin ^{2}\left( \tfrac{\pi d_{2}}{L_{m}}\right) \right]%
\end{array}%
\end{equation}
here $\theta ^{m}$\ and \textit{L}$_{m}$\ are the mixing angle and the
oscillation length inside the Earth respectively, and $d_{i}$\ is the
distance travelled by neutrinos inside the Earth before reaching the
detector $D_{i}$.

$\qquad $\textit{2)} The second factors is a summing of two-factor terms.
Let us try to see how they behave:

$^{\ast }$ $\left( F_{e}^{o}-F_{x}^{o}\right) $ and $\left( F_{\bar{e}%
}^{o}-F_{x}^{o}\right) :$ Since the $\nu _{e}$ ($\bar{\nu}_{e}$) spectrum is
softer than the $\nu _{x}$ spectrum, and the luminosities of both the
spectra are similar in magnitude \cite{Janka2}, the term ($%
F_{e}^{o}-F_{x}^{o}$), (and therefore $\left( F_{\bar{e}}^{o}-F_{x}^{o}%
\right) $), is positive at \textit{LOW} energies and becomes negative at
higher energies where the $\nu _{x}$ flux overwhelms the $\nu _{e}$ ($\bar{%
\nu}_{e}$) flux. Therefore, the Earth effect has a different sign for 
\textit{LOW} and high energies, and there exists a critical energy $E_{c}$ ($%
\bar{E}_{c}$), such that $F_{e}^{o}$($E_{c}$)$=F_{x}^{o}$($E_{c}$), ($F_{%
\bar{e}}^{o}$($\bar{E}_{c}$)$=F_{x}^{o}$($\bar{E}_{c}$)), where this change
of sign takes place. Since the cross section of the neutrino interactions
increases with energy, the Earth effect is expected to be more significant
at higher energies (if all the other factors are only weakly sensitive to
the neutrino energy).

$^{\ast }$ $\left( a_{2e}-a_{1e}\right) $, $\left( a_{\bar{2}e}-a_{\bar{1}%
e}\right) $, $\left( a_{2\bar{e}}-a_{1\bar{e}}\right) $ or $\left( a_{\bar{2}%
\bar{e}}-a_{\bar{1}\bar{e}}\right) $:\ They represent the differences
between the $a_{il}$ probabilities and can be extracted from Eq. (\ref{1}),
Eq. (\ref{2}), Eq. (\ref{3}) and Eq. (\ref{4}). They are seen to be
dependent on the mass hierarchy as well as the specific solar neutrino
solution (\textit{LMA} or \textit{LOW}). Non vanishing values for this
difference make the observation of the Earth matter effect possible.


\begin{thebibliography}{99}
\bibitem{SN1987} Hirata. K.S et. al., Kamiokande collaboration, Phys. Rev.
Lett. 58 (1987) 1490. Bionta. R.M et. al., IMB collaboration, Phys. Rev.
Lett. 58 (1987) 1494.

\bibitem{SK} Fukuda. Y et al., Phys. Rev. Lett. 82 (1999) 2644, K. Scholberg
(for the SuperKamiokande collaboration), hep-ex/9905016.

\bibitem{SNO} The SNO collaboration, Nucl. Instrum. Meth. A449 (2000)
172-207, nucl-ex/9910016.

\bibitem{Macro} Ambrosio. M et. al., MACRO Collaboration, Astropart. Phys. 8
(1998) 123-133.

\bibitem{Solarnu} Bahcall. J.N, Krastev. P.I and Smirnov. A.Yu, Phys. Rev D
58 (1998) 096016. Krastev. P.I, hep-ph/9905458. Gonzalez-Garcia. M.C,
deHolanda. P.C, Pena-Garayand. C and Valle. J.W.F, Nucl. Phys. B573 (2000)
3-26, hep-ph/9906469.

\bibitem{MSW} Mikheyev. S.P and Smirnov. A.Yu, Sov. J. Nucl. Phys. 42 913-17
(1985), Nuovo Cimento 9 C 17-26 (1986), Wolfenstein. L, Phys. Rev. D 17
(1978) 2369-74.

\bibitem{RSFP} Fujikawa. K and Shrock. R.E, Phys. Rev. Lett 45 (1980) 963;
Lim. C.-S and Marciano. W.J, Phys. Rev D 37 (1988) 1368.

\bibitem{magnetic} Particle Data Group, European Physical Journal C 15
(2001) 1, hep-ex/9907037.

\bibitem{RSFP+SN} Nunokawa. H, Qian. Y.-Z and Fuller. G.M,
astro-ph/96100209. Akhmedov. E.Kh, Lanza. A, Petcov. S.T and Sciama. D.W,
hep-ph/9603443. Athar. H, Peltoniemi. J.T and Smirnov. A.Yu, Phys. Rev. D
51(1995) 6647, hep-ph/9501283. Athar. H, hep-ph/9902222.

\bibitem{Nucl} McLaughlin. G.C, Fetter. J.M, Balantekin. A.B and Fuller.
G.M, Phys. Rev. C 59 (1999) 2873-2887, astro-ph/9902106.

\bibitem{DigheSmirnov} Dighe. A.S and Smirnov. A.Yu, Phys. Rev D 62 (2000)
033007, hep-ph/9907423.

\bibitem{Kamland} KamLAND collaboration, Phys. Rev. Lett. 90 (2003) 021802,
hep-ex/0212021.

\bibitem{Bethe} Bethe. H.A, Rev. Mod. Phys. 62, 802 (1990).

\bibitem{Suzuki} Suzuki. H, ''Supernova Neutrinos'', Physics and
Astrophysics of Neutrinos, eds. Fukugita. M and Suzuki. A, Springer Verlag,
1994, pp 763.

\bibitem{BetheWilson} Bethe. H.A and Wilson. J.R, ApJ 295 (1985) 14.

\bibitem{WoosWea} Woosley. S.E and Weaver. T.A, Annu. Rev. Astron.
Astrophys. 24, 205 (1986).

\bibitem{density} Janka. H.-T and Hillenbrandt. W, Astron. Astrophys. Suppl.
78 (1989) 375. Notzold. D, Phys. Lett. B 196 (1987) 315.

\bibitem{Maki} Maki. Z, Nakagawa. M and Sakata. S, Theor. Phys. 28 (1962)
870.

\bibitem{Botella} Botella. F.J, Lim. C.-S and Marciano. W.J, Phys. Rev. D 35
(1987) 896.

\bibitem{MagMoment} Derbin. A.I et al., JETP Lett.57, 768 (1993).

\bibitem{Chooz} Apollonio. M et. al., Phys. Lett. B466 (1999) 415-430, the
CHOOZ collaboration, hep-ex/9907037.

\bibitem{Smirnov} Mikheyev. S.P and Smirnov. A.Yu, Sov. J. Nucl. Phys. 42
(1985) 913.

\bibitem{Landau} Landau. L, Phys. Z. Sowjetunion 2 (1932) 46. Zener. C,
Proc. R. Soc. London, Ser. A 137 (1932) 696.

\bibitem{Akhmedov} Akhmedov. E.Kh, Sov. Phys. JETP 68 (1989) 690.

\bibitem{LVD} Aglietta. G et. al., LVD Collaboration, Nuovo Cim. 18 C (1995)
629.

\bibitem{Raffelt} Raffelt. G.G, ''Stars as Laboratories for Fundamental
Physics'', The University of Chicago Press, 1996.

\bibitem{Janka1} Janka. H.-T and Hillenbrandt. W, Astron. Astrophys. Suppl.
78 (1989) 375.

\bibitem{LunardiniSmirnov} Lunardini. C, Smirnov. A.Yu, Nucl. Phys. B 616
(2001) 307-348, hep-ph/0106149. Phys. Rev. D 63 (2001) 073009,
hep-ph/0009356.

\bibitem{Janka2} Janka. H.-T, Astropart. Phys. 3 (1995) 377-384,
astro-ph/9503068.
\end{thebibliography}
\end{document}